\documentclass[pdftex,aps,prl,twocolumn,groupedaddress,superscriptaddress,amsfonts,amssymb,amsmath,citeautoscript,a4paper]{revtex4-1}

\usepackage{orcidlink}
\usepackage{lipsum}
\usepackage[utf8]{inputenc}
\usepackage[english]{babel}
\usepackage{microtype}
\usepackage{txfonts}
\usepackage{txfontsb}
\usepackage{upgreek}
\usepackage{bm}
\usepackage{graphicx}
\usepackage{float}
\usepackage{xspace}
\usepackage[skip=0pt, indent=15pt]{parskip}
\usepackage{titlesec}
\titleformat{\paragraph}[runin]{\normalfont\normalsize\bfseries}{}{0}{}[]
\titleformat{\section}[hang]{\normalfont\large\bfseries}{}{0}{}[]
\usepackage{subfiles}
\usepackage{enumerate}
\usepackage[inline]{enumitem}

\usepackage{siunitx}
\DeclareSIUnit[number-unit-product=]\percent{\char`\%} % remove space before percentage "units"

\usepackage{hyperref}
\hypersetup{colorlinks,
            linkcolor={blue!75!black!80!yellow},
            citecolor={blue!75!black!80!yellow},
            urlcolor={blue!75!black!80!yellow},
            pdfstartview=FitH}

\makeatletter
\renewcommand{\fnum@figure}{FIG.~\thefigure}
\makeatother

\usepackage{mdframed}
\definecolor{blue-violet}{rgb}{0.54, 0.17, 0.89}

\newmdenv[topline=false, rightline=false, bottomline=false,%
  linewidth=1.25pt, innerrightmargin=0pt, leftmargin=-8pt,%
  innerleftmargin=7pt, skipabove=0pt, skipbelow=0pt,%
  linecolor=blue-violet, fontcolor=blue-violet]{mdleftbar}

\begin{document}

\title{Near-field refractometry of van der Waals crystals}

\author{Martin Nørgaard\,\orcidlink{0009-0002-5412-4669}}
\affiliation{POLIMA---Center for Polariton-driven Light--Matter Interactions, University of Southern Denmark, Campusvej 55, DK-5230 Odense M, Denmark}

\author{Torgom Yezekyan\,\orcidlink{0000-0003-2019-2225}}
\affiliation{POLIMA---Center for Polariton-driven Light--Matter Interactions, University of Southern Denmark, Campusvej 55, DK-5230 Odense M, Denmark}

\author{Stefan Rolfs\,\orcidlink{0009-0007-8489-1013}}
\affiliation{Center for Nano Optics, University of Southern Denmark, Campusvej 55, DK-5230~Odense~M, Denmark}

\author{Christian~Frydendahl\,\orcidlink{0000-0003-4734-9923}}
\affiliation{Center for Nano Optics, University of Southern Denmark, Campusvej 55, DK-5230~Odense~M, Denmark}
\affiliation{Department of Physics and Astronomy, Aarhus University, Ny Munkegade 120, DK-8000~Aarhus~C, Denmark}

\author{N. Asger Mortensen\,\orcidlink{0000-0001-7936-6264}}
\affiliation{POLIMA---Center for Polariton-driven Light--Matter Interactions, University of Southern Denmark, Campusvej 55, DK-5230 Odense M, Denmark}
\affiliation{Danish Institute for Advanced Study, University of Southern Denmark, Campusvej 55, DK-5230 Odense M, Denmark}

\author{Vladimir A. Zenin\,\orcidlink{0000-0001-5512-8288}}
\affiliation{Center for Nano Optics, University of Southern Denmark, Campusvej 55, DK-5230~Odense~M, Denmark}

\email{zenin@mci.sdu.dk; asger@mailaps.org}

\date{\today}

\begin{abstract}
Common techniques for measuring refractive indices, such as ellipsometry and goniometry, are ineffective for van der Waals crystal flakes because of their high anisotropy and small, micron-scale, lateral size. To address this, we employ near-field optical microscopy to analyze the guided optical modes within these crystals. By probing these modes in MoS$_2$ flakes with subwavelength spatial resolution at a wavelength of $1570\,\mathrm{nm}$, we determine both the in-plane and out-of-plane permittivity components of MoS$_2$ as $16.11$ and $6.25$, respectively, with a relative uncertainty below $1\%$, while overcoming the limitations of traditional methods.

\end{abstract}

\maketitle

\section{Introduction}

The discovery of the exceptional properties of graphene, enabled by a straightforward exfoliation technique to isolate monocrystalline graphitic films in the early 2000s~\cite{Novoselov:2004}, ignited a search for other materials with layered structures held together by weak van der Waals (vdW) forces. These materials, including transition metal dichalcogenides (TMDCs), have become a focal point in optoelectronics and quantum nanophotonics owing to their unique optical and electrical properties~\cite{Song:2024,Goncalves:2020,Reserbat-Plantey:2021,Zhang:2021}. 
Due to the layered structure of vdW materials, most of them are expected to exhibit giant anisotropy and even hyperbolic dispersion~\cite{Gjerding:2017}, which cannot be found among previously known naturally occurring materials.
TMDCs, for instance, display remarkable optical behavior; their monolayers, readily exfoliated from bulk crystals~\cite{Novoselov:2005,Velicky:2017,Frisenda:2020}, often exhibit optoelectronic properties distinctly different from those of their bulk counterparts~\cite{Liu:2016,Rasmussen:2015}. Molybdenum disulfide (MoS$_2$), the TMDC considered in the present work, exemplifies this: it transitions from an indirect bandgap semiconductor with a 1.29\,eV bandgap in bulk~\cite{Boker:2001} to a direct bandgap semiconductor with a 1.8--1.9\,eV bandgap as a monolayer~\cite{Mak:2010,Li:2014}.

Theoretical and computational advancements~\cite{Cappelluti:2013,Kormanyos:2015,Fang:2015,Rasmussen:2015,Thygesen:2017} and experimental explorations~\cite{Novoselov:2005,Tang:2014,Shen:2013,Li:2014,Yim:2014} have propelled the understanding of vdW materials, yet the experimental characterization of their basic optical properties remains challenging. 
Exfoliated flakes are generally too small and non-uniform in thickness for traditional refractive index measurements, which limits precision in determining key optical parameters like the complex permittivity tensor.
Understanding these optical properties is crucial for the design and optimization of devices based on vdW materials, such as metasurfaces~\cite{Verre:2019,Munkhbat:2020,Tonkaev:2024}, and could enable new applications previously limited by a lack of suitable materials with such unique properties.

Some refractive index measurement methods rely on Snell's law and goniometer setups to precisely measure refraction angles; however, they are ineffective for slab-like samples, where refraction results only in slight beam displacement. Techniques such as measuring critical angles for total internal reflection or using Brewster's angle provide rough estimations but lack accuracy due to weak angular dependence. The most advanced method, ellipsometry, measures the change in polarization upon reflection, which is fitted by different models to extract the thickness and refractive index of the unknown layer. However, the precision of conventional spectroscopic ellipsometry relies on using a well-collimated beam with a spot size of $>300\,\mathrm{\upmu m}$, which is too large to make it reliable for the investigation of exfoliated flakes, because defects and non-uniformities are also illuminated~\cite{Ermolaev:2020,Munkhbat:2022}. This limitation is overcome in an imaging ellipsometry setup, which uses angled wide-field illumination and microscopy detection with an objective, claimed to have a lateral resolution below $10\,\mathrm{\upmu m}$~\cite{Ermolaev:2021,Funke:2016,Zotev:2023,Vyshnevyy:2023,Grudinin:2023}. However, its primary purpose is imaging of material contrast, which might compromise the accuracy of refractive index measurements, especially for challenging samples as anisotropic TMDC flakes. Finally, the large refractive index of most TMDC materials results in low sensitivity of the above far-field methods to the out-of-plane component of the refractive index, because the incident illumination from the air even at large angles will be refracted to nearly normal direction inside the flake.

\begin{figure*}[htb!]
    \centering
\includegraphics[]{ 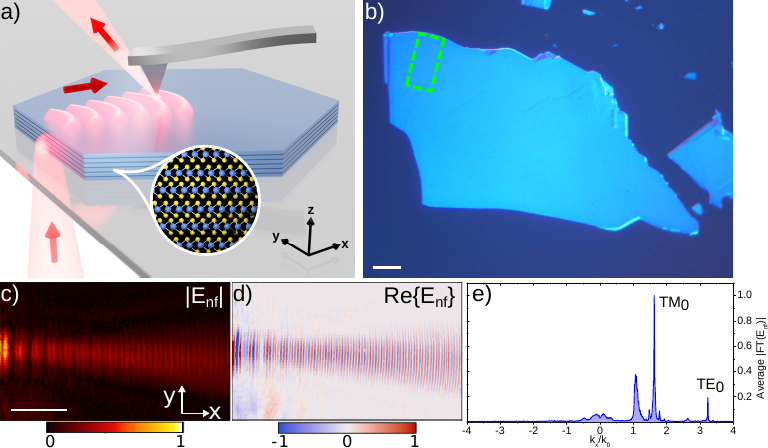}
    \caption{\textbf{a)}~Artistic representation of the experiment where guided modes in an MoS$_2$ flake are probed using a transmission s-SNOM setup. \textbf{b)}~Differential interference contrast image of a $t=185$\,nm thick MoS$_2$ flake. The scale bar is 20\,$\mathrm{\upmu m}$. \textbf{c-d)}~Amplitude $\left|E_\mathrm{nf}\right|$ and real part $\mathrm{Re}\left\{E_\mathrm{nf}\right\}$ of the complex near-field map of the marked $40\,\mathrm{\upmu m}$ by $20\,\mathrm{\upmu m}$ region in panel~b) (green dashed rectangle). The scale bar is 10\,$\mathrm{\upmu m}$. \textbf{e)}~Momentum-space representation of the complex near-field map, highlighting Fourier components associated with TE$_0$ and TM$_0$ modes.}  
    \label{fig:fig1}
\end{figure*}

In contrast to far-field methods, scanning near-field microscopy (SNOM) bypasses the issue of lateral resolution.
It has been shown that one can use scattering-type (s-)SNOM to obtain quantitative measurements of local dielectric constants by the modification of the scattering strength of the s-SNOM probe by its environment. One approach is based on developing a rigorous model of the probe (a point dipole model, which was later modified into a finite dipole model), which is first applied on known samples for calibration, and then it can be used for measurements~\cite{Govyadinov:2013,Knoll:2000}.  Another approach is based on an empirical search for the probe response function from a set of calibration measurements (black-box calibration)~\cite{Guo:2021,Siebenkotten:2024}.  These methods allow for the extraction of permittivity without detailed electromagnetic modeling of the probe-sample interaction, even accounting for probe tapping effects and far-field background. However, they are developed for isotropic samples, making it challenging to apply them to highly anisotropic vdW materials. Moreover, to our knowledge, they cannot deal with samples where guided modes are excited.

Here we implemented a method we termed `near-field refractometry', which works by probing guided modes within the material, whose properties are directly linked to the material optical properties~\cite{Boroviks:2021,Lebsir:2022,Casses:2022,Casses:2024,Jang:2024}. Moreover, our approach demonstrated sensitivity to both in-plane and out-of-plane refractive index components, making it uniquely suited for measurements of highly anisotropic vdW materials. 

We demonstrate the precision and accuracy of our method by investigating a highly anisotropic vdW material, namely MoS$_2$~\cite{Ermolaev:2021}, using a phase-resolved SNOM (Figure~\ref{fig:fig1}a).
We map the near field of guided modes within MoS$_2$ flakes of finite thickness (Figure~\ref{fig:fig1}c and d) at photon energies below the bandgap ($\lambda_0 = 1570$\,nm), with a scan size as small as $40\,\mathrm{\upmu m}$ by $20\,\mathrm{\upmu m}$ (Figure~\ref{fig:fig1}b). First, the guided transverse electric (TE) and transverse magnetic (TM) modes are found by Fourier transforming the recorded near-field map (Figure~\ref{fig:fig1}e). After filtering in the Fourier domain, these modes are fitted in the direct space to extract their propagation constant. Finally, after collection of the thickness-dependent dispersion characteristics of these modes, we extract the anisotropic dielectric function of MoS$_2$ and estimate its uncertainty.

\section{Results}
To illustrate the general concept of near-field refractometry of vdW materials, we first provide the electrodynamic theoretical foundation, followed by sample fabrication, near-field optical measurements, and data processing to eventually extract the anisotropic optical constants of the vdW material.

\begin{figure}[t!]
    \centering
\includegraphics[width=\columnwidth]{ 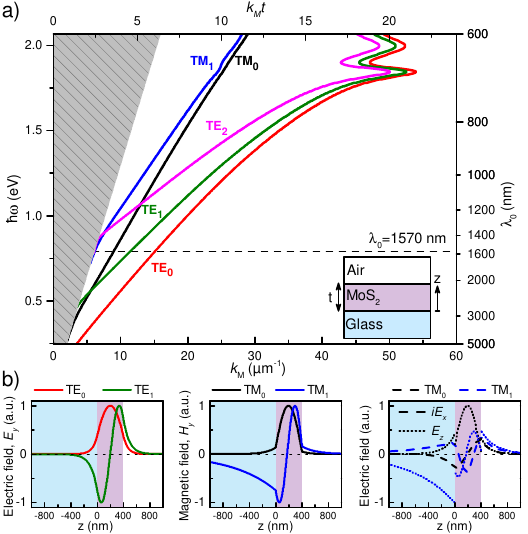}
    \caption{~\textbf{a)}~Dispersion diagram showing photon energy $\hbar\omega$ (left vertical axis) and corresponding free-space wavelength $\lambda_0$ (right vertical axis) versus mode propagation constant $k_M$ (horizontal axes) for a $t=400$\,nm MoS$_2$ flake on BK7 glass substrate for selected TE$_m$ ($m=0, 1, 2$) and TM$_m$ ($m=0, 1$) modes. The regime of leaky substrate modes is gray shaded, while the horizontal dashed line indicates the experimentally used wavelength.
    \textbf{b)}~Field profiles for the TE$_m$ and TM$_m$ modes ($m=0, 1$), propagating along the $x$-axis.}  \label{fig:fig2}
\end{figure}

\textbf{Planar thin film waveguide modes.} 
We consider a generic anisotropic dispersive dielectric function of a uniaxial vdW crystal flake
\begin{equation}\label{eq:MoS2}
\varepsilon(\omega)=\begin{pmatrix} \varepsilon_\parallel(\omega) & 0 & 0\\
    0 & \varepsilon_\parallel(\omega) & 0\\
    0& 0 & \varepsilon_\perp(\omega)
    \end{pmatrix},
\end{equation}
where $\varepsilon_\parallel$ and $\varepsilon_\perp$ are the in-plane and out-of-plane components, respectively. We assume that the flake of thickness $t$ is resting on a substrate with dielectric function $\varepsilon_s(\omega)$, while being exposed to air above the flake ($\varepsilon_a = 1$). Given its high dielectric function ($\varepsilon>\varepsilon_s>\varepsilon_a$), the flake is essentially a thin-film waveguide, supporting TE and TM modes, which are guided in-plane of the flake, along the $x$-axis, while being strongly localized in the out-of-plane direction ($z$-direction). The dispersion of TE modes is governed by
\begin{subequations}
\begin{equation}
q_\mathrm{TE} t = \mathrm{atan}\left(\frac{k_{a}}{q_\mathrm{TE}}\right) + \mathrm{atan}\left(\frac{k_{s}}{q_\mathrm{TE}}\right) + m\pi,
\label{eq:disp_TE}
\end{equation}
where $k_{a,s} = \sqrt{k_M^2-\varepsilon_{a,s} k_0^2}$ are the confinement factors in the air and the substrate, respectively, and $q_\mathrm{TE} = \sqrt{\varepsilon_\parallel k_0^2 - k_M^2}$, while $k_M$ is the mode propagation constant, $k_0=\omega/c$ is the free-space wave vector, and $m$ is an integer associated with the mode order. Likewise, the TM modes are governed by
\begin{equation}
q_\mathrm{TM} t = 
\mathrm{atan}\left(\frac{k_{a}\varepsilon_\parallel}{q_\mathrm{TM}\varepsilon_a}\right) + 
\mathrm{atan}\left(\frac{k_{s}\varepsilon_\parallel}{q_\mathrm{TM}\varepsilon_s}\right) + m\pi,
\label{eq:disp_TM}
\end{equation}
\label{eq:disp}
\end{subequations}
where $q_\mathrm{TM} = \sqrt{\varepsilon_\parallel/\varepsilon_\perp}\sqrt{\varepsilon_\perp k_0^2 - k_M^2}$. The complex-valued dispersion relation $k_M(\omega)=k_M'(\omega)+ik_M''(\omega)$ for TE$_m$ and TM$_m$ modes of any mode order $m$ can be obtained by numerically solving Equations~\ref{eq:disp_TE} and \ref{eq:disp_TM} for a given frequency $\omega$, flake thickness $t$, and dielectric functions $\varepsilon(\omega)$ and $\varepsilon_s(\omega)$. The solution for a $t=400$\,nm thick MoS$_2$ flake, supported by a BK7 glass substrate, results in the dispersion diagram illustrated in Figure~\ref{fig:fig2}a, where we have used experimentally tabulated dispersive parameters for the MoS$_2$ in the Tauc--Lorentz model for $\varepsilon(\omega)$~\cite{Ermolaev:2021}, while for the BK7 glass, $\varepsilon_s(\omega)$ is conveniently represented by the Sellmeier dispersion formula. For more details on these representations, see the Supplementary Information (SI) section S1. In the dispersion diagram, the regime of leaky substrate modes ($k_M < \sqrt{\varepsilon_s}\omega/c$) is gray shaded. 

At a specific photon energy $\hbar\omega$ (or wavelength $\lambda_0 = 2\pi/k_0= 2\pi c/\omega$), only certain guided modes are supported by the MoS$_2$ flake. Adjusting the thickness $t$ shifts the guided mode curves along the $k_M = \sqrt{\varepsilon_s} \omega/c$ line. Consequently, increasing $t$ permits more and higher-order modes, while decreasing $t$ limits the number of allowed modes in the MoS$_2$ flake. 
In addition to the characteristic thin-film waveguide dispersion relation at low energies, one can also note a clear hybridization with excitons at photon energies around $\hbar\omega \sim$ 1.8--2\,eV.

Importantly, SNOM maps the evanescent field in the vicinity of the probe, therefore the mode confinement plays a crucial role in detecting the individual modes. We have calculated the mode profiles at $\lambda_0=1570\,\mathrm{nm}$, which are depicted in Figure~\ref{fig:fig2}b.
Since s-SNOM is most sensitive to the out-of-plane component of the electric field due to its elongated tip-based scattering probe, the signal of the TM modes will be the strongest, as they contain both $E_x$ and $E_z$ electric field components. In contrast, TE modes only have the $E_y$ field component. More details on the influence of the modes on the detectability is presented in the SI section S2. 

Our proposed near-field refractometry method starts with measuring the complex-valued near-field maps of the guided modes, allowing accurate determination of the propagation constant $ k_M(t)$ of each mode, which is then rigorously fitted to the above model, Equations~\ref{eq:disp_TE} and \ref{eq:disp_TM}, to eventually extract $\varepsilon_\parallel$ and $\varepsilon_\perp$.

\textbf{Sample fabrication and pre-selection of flakes.} 
We mechanically exfoliated MoS$_2$ flakes with a modified 'scotch tape' method (see Methods section).
Figure~\ref{fig:fig1}b shows a differential interference contrast image of one flake, revealing surface details not clearly visible in the bright- and dark-field images. However, all three complementary imaging methods were used to select flakes with a clean and uniformly thick area, suitable for the precise s-SNOM scanning.
Another important aspect of the exfoliation technique used is that as MoS$_2$ flakes become thicker, they tend to exhibit more folds and wrinkles, making it harder to obtain large, uniform, flat areas.

It is crucial to have a wide range of flake thicknesses, which should support several guided modes to reliably fit the data to Equations~\ref{eq:disp}a-b and verify our model, as will be discussed later.
For this purpose, we selected six different flakes with the thicknesses varied from ca. 80 to 460\,nm, as measured with atomic force microscopy (AFM), each with an estimated uncertainty of $\pm 10\%$.

Furthermore, reflection spectroscopy of each flake can be used to estimate the thickness using the Fresnel equations for anisotropic media~\cite{Hsu:2019}; however, this requires \emph{a priori} knowledge of $\varepsilon(\omega)$ over a wide range of frequencies.

For a detailed view of the MoS$_2$ flake characterization, including optical microscopy and reflection spectroscopy, see the SI section S3.

\textbf{Near-field measurements.} 
Our s-SNOM setup can measure both the amplitude and the phase of the evanescent near field, enabling the complete mapping of the complex dispersion relation of guided modes. In our s-SNOM configuration, the light reaches the sample from the opposite side of the scattering tip, which is referred to as transmission-type~\cite{Boroviks:2021,Lebsir:2022}. Unlike the commonly used reflection configuration~\cite{Hu:2017}, this transmission configuration conveniently separates the illumination and detection parts of the setup. This allows mapping the near field of guided modes 'as launched' from the edge of the MoS$_2$ flakes, without taking the influence of the tip and geometrical decay into account. 
The concept of transmission s-SNOM is illustrated in Figure~\ref{fig:fig1}a, while Figure~\ref{fig:fig1}c and d present an example of the obtained near-field map, displayed as the electric near-field amplitude $\left|E_\mathrm{nf}\right|$ and its real part $\mathrm{Re}\{E_\mathrm{nf}\}$ for the $t=185$\,nm thick MoS$_2$ flake. 
The characteristic fringes in the near-field maps indicate the excitation of more than one guided mode and the interference with the 'background' signal.   
A more detailed schematic of the transmission type s-SNOM is presented in the SI section S4.

In the experiments, a continuous-wave laser with a wavelength of $\lambda_0 = 1570$\,nm and a Gaussian beam profile is used as the light source. In this wavelength region, MoS$_2$ is expected to have negligible optical loss~\cite{Ermolaev:2021}, meaning that the drop in $\left|E_{\mathrm{nf}}\right|$ along the propagation is due to the divergence of the beam itself (Figure~\ref{fig:fig1}c). As an alternative to the dispersion diagram, where the frequency is varied (Figure~\ref{fig:fig2}a), we instead vary the flake thickness in the experiment. Also, we represent the mode propagation constant in terms of the effective mode index, $N_m = k_M/k_0$, which makes a direct link with the material properties. Essentially, $N_m$ is the weighted average refractive index, experienced by the mode, where the $E$-field of the mode profile is used as the weight. Therefore, the upper limit for $N_m$ is $\varepsilon$, when nearly all of the mode $E$-field is concentrated inside flake, whereas $N_m\approx\sqrt{\varepsilon_s}$ indicates that the mode is weakly confined (close to be leaky substrate mode).

For each scan, we choose an area that encompasses the entire beam along the $y$-axis and is sufficiently large along the $x$-axis to capture enough periods, resulting in a scan size of approximately $40\,\rm{\upmu m}$ by $20\,\rm{\upmu m}$ for all measurements. 
To avoid artifacts (aliasing) and distortion in the complex near-field data, the spatial sampling rate must exceed the Nyquist rate ($f_s>2f_\mathrm{max}$) for the highest spatial frequency ($f_\mathrm{max}=k_{\parallel}/2\pi$) 
measured along the propagation direction.
To ensure that we resolve all modes for all thicknesses while maintaining consistency, a step size of 25\,nm along the propagation direction is chosen for all measurements. Along the $y$-axis, we used a step size of 250\,nm.

\begin{figure}[t!]
    \centering
    \includegraphics[width=\linewidth]{ 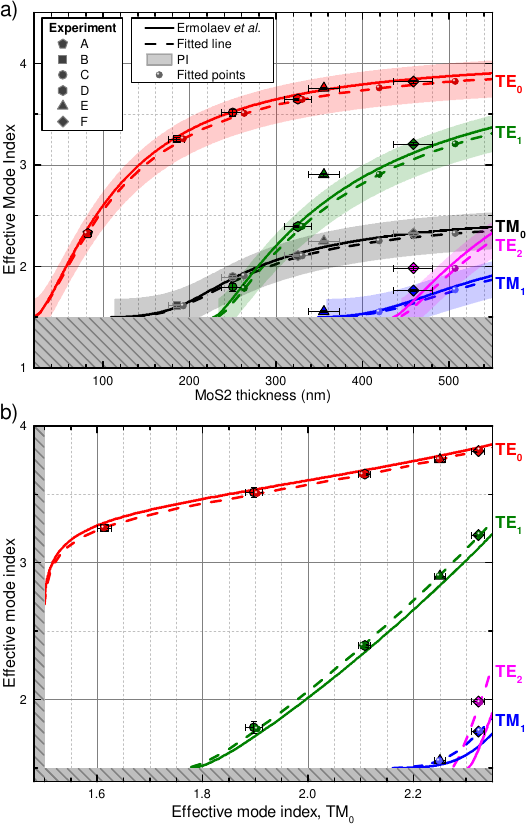}
    \caption{\textbf{a)}~The effective mode index $N_m$ of the different TE$_m$ ($m=0, 1, 2$) and TM$_m$ ($m=0, 1$) modes as a function of thickness $t$ of MoS$_2$. \textbf{b)}~Parametric plot of the effective mode index of different TE$_m$ ($m=0,1,2$) and TM$_m$ ($m=1$) modes compared to the TM$_0$ mode with the thickness $t$ varied parametrically. The regimes of leaky substrate modes are gray shaded in both panels. 
    Data points with error bars represent our measurements for flakes A through F, while curves are derived from the dispersion relations using reference permittivities from Ermolaev~\emph{et~al.}~\cite{Ermolaev:2021} (solid lines) and our fitted permittivities (dashed lines), which also contain points corresponding to the investigated flakes. 
    Additionally, the shaded areas along the curves in panel~a) represent the $95\%$ prediction interval (PI) for the fitted lines. }  
    \label{fig:fig3}
\end{figure}

\textbf{Data processing.}
To unambiguously extract the 'pure' or individual guided modes, the complex near-field maps are filtered to consider only one guided mode at a time. First, by applying a one-dimensional (1D) Fourier transform along the propagation direction ($x\rightarrow k_x$) and averaging along the columns ($y$-axis), we obtain an overview of guided modes supported by the flake.
An example of this is shown in Figure~\ref{fig:fig1}e, which presents a spectrum containing three prominent peaks. One peak is located close to $k_x/k_0 \approx 1$, corresponding to the refractive index of free space; this peak we attribute to the diffraction of the incident light from the edge of the flake and other reflections in the system. The second peak, around $k_x/k_0 \approx 1.6$, is associated with the TM$_0$ mode, and is the strongest because of the high sensitivity of s-SNOM to the out-of-plane component of the $E$-field. The third peak, around $k_x/k_0 \approx 3.25$, corresponds to the TE$_0$ mode. In general, these labels can be assigned by solving Equations~\ref{eq:disp_TE} and~\ref{eq:disp_TM} and using a suitable initial guess for the flake dielectric constant~\cite{Ermolaev:2021}. Alternatively, if optical properties are unknown, the mode assignment can be done by polarizing the normally incident beam along or across the flake edge, which will excite predominantly TE or TM modes, correspondingly. The Fourier spectrum in Figure~\ref{fig:fig1}e also shows relatively low signal around $k_x \sim 0$, indicating the effectiveness of background removal, while the absence of the signal at $k_x < 0$ suggests very low back-reflection of the guided modes ($k_x \rightarrow -k_x$).

One can directly determine the mode propagation constant $k_M$ from the Fourier spectrum, however, the accuracy will be low due to the limited size of the scan. Additionally, since the measured near field does not drop to zero at the edges of the scan, its Fourier spectrum will contain artifacts from the windowing function (also known as apodization function).
The commonly used zero padding artificially improves the Fourier resolution, but still plagues the spectrum with artifacts from applying the window, causing spectral leakage of modes in the Fourier domain.
To overcome these limitations, we apply extended discrete Fourier transform (EDFT), which essentially iteratively extrapolates the limited-range data to match its Fourier spectrum with the one of the infinite-range data, assuming that its Fourier spectrum is band limited~\cite{liepins:EDFT}. Finally, to rely less on the parameters of the Fourier transform, we filter each mode in the Fourier domain and fit it in the direct space to determine its propagation constant.

First, we inspect that the field of the filtered mode, $E_m$, follows the expected distribution of the 2D Gaussian beam, propagating along the $x$-axis, which can be written in the following \emph{complex} form~\cite{Arnaud:1985}:
\begin{equation}
    E_m(x,y)=E_0\frac{w_0}{w}\exp{\left(-\frac{y^2}{w^2}+iN_mk_0x\right)},
\end{equation}
where $w=w_0\sqrt{1+ix/x_R}$ is the \emph{complex-valued} diverging beam waist ($w_0\approx3\,\upmu\mathrm{m}$ is roughly equal to the waist of the incident beam, focused on the flake edge) and $x_R=\tfrac{1}{2} N_mk_0 w_0^2$, is the Rayleigh length of the beam. To compensate the divergence, we integrate $E_m (x,y)$ in the $y$-direction, which should result in a pure exponential dependence $\exp(iN_m k_0 x)$ for the ideal 2D Gaussian beam. Since we don't expect any absorption losses for the chosen wavelength (meaning $N_m$ is a real number), we find $N_m$ from a slope of the unwrapped phase of $\int E_m(x,y)\,dy$, with an associated standard error $\Delta$. 

The above procedure is repeated for all the measurements for the same flake. Finally, we average $N_m$ within this set using $1/\Delta^2$ as a weight, and estimate its uncertainty, accounting for both the error of individual point and the variance of $N_m$ within the set:

\begin{equation}  \Delta{N}_{\mathrm{eff}}=\sqrt{\mathrm{avg}_w(\Delta^2)+\mathrm{var}_w(N_{\mathrm{eff}})},
\label{eq:delta}
\end{equation}
where $w$ indicates the same weight $1/\Delta^2$. Technical details about the processing of the near-field data can be found in the SI section S5.
The obtained effective mode indices for each flake thickness and their associated errors are summarized in Figure~\ref{fig:fig3}, which also shows the numerical solutions for each mode corresponding to Equations~\ref{eq:disp_TE} and \ref{eq:disp_TM} for varying thickness $t$.

\textbf{Experimental data fitting. }
Given the effective mode indices for each mode at different thicknesses $t$, it is possible to estimate $\varepsilon_\parallel$ and $\varepsilon_\perp$ for MoS$_2$ at the used wavelength (here $\lambda_0=1570$\,nm). The dispersion relations implicitly defines the effective mode indices, which we denote as $N_{m}(\varepsilon,t)$ for ease of notation, as functions of $t$ and $\varepsilon$ (which includes the two components $\varepsilon_\parallel$ and $\varepsilon_\perp$). Therefore, we only treat $t^\mathrm{fit}$ and $\varepsilon^\mathrm{fit}$ as free fitting parameters, and fit our measurements by minimizing the following squared errors 
\begin{equation}
   F_{i}(\varepsilon^{\mathrm{fit}},t^\mathrm{fit}_i)=\left(\frac{t^\mathrm{fit}_i-t_i}{\Delta t_i}\right)^2+\sum_m\left(\frac{N_m(\varepsilon^{\mathrm{fit}},t^\mathrm{fit}_i)-N_{i,m}}{\Delta N_{i,m}}\right)^2,
   \label{eq:mse}
\end{equation}
where $i$ runs over the number of different MoS$_2$ flake thicknesses and $m$ runs over the different TE and TM modes. The estimated errors of the measurements ($\Delta t_i$ and $\Delta N_{i,m}$) are used as  weights to normalize the discrepancy between measured and fitted $t$ and $N$, accordingly. As such, the measurements with a large error will have small influence on the fitting.

One way to perform the fitting is to search for all 8 free fitting parameters at once (two for $\varepsilon^\mathrm{fit}$ and six for $t^\mathrm{fit}$ corresponding to the six different flakes). However, this demands high computational resources and there is a risk of ending in a local minimum. Instead, we used the following nested solver structure. If $\varepsilon$ is known, then the thickness of each flake can be determined separately by minimizing Equation~\ref{eq:mse} for each flake, which essentially makes implicit definition of $t^\mathrm{fit} = t^\mathrm{fit}\left(\varepsilon\right)$. Then this function (involving least-squares solver) is inserted into Equation~\ref{eq:mse}, which is transformed to contain only the free fitting parameter of $\varepsilon^\mathrm{fit}$. The latter can easily be found by another least-squares solver.

To find the uncertainty of the determined $\varepsilon^\mathrm{fit}$, we apply a perturbation approach, where we modify one of the input parameters ($t_i$ or $N_{i,m}$) and re-iterate the fitting to estimate the sensitivity of $\varepsilon^\mathrm{fit}$ to each parameter. We then estimate the uncertainty as the following:
\begin{equation}
    \Delta\varepsilon^{\mathrm{fit}}=\sqrt{\sum_i\left[\left(\frac{\partial\varepsilon^{\mathrm{fit}}}{\partial t_i}\Delta t_i\right)^2+\sum_m\left(\frac{\partial \varepsilon^{\mathrm{fit}}}{\partial N_{i,m}}\Delta N_{i,m}\right)^2\right]}.
\end{equation}
Further details on the fitting procedure are presented in the SI section S6.

Utilizing this method yields a set of fitted thicknesses for each flake, as summarized in table~\ref{tab:t_fit}, and the two permittivity components and their associated error, as seen in table~\ref{tab:epsilon_fit}.

\begin{table}[!h]
\centering
\begin{tabular}{l|l|l|l|l|l|l}
Flake       & A     & B      & C      & D      & E      & F      \\ \hline
$t$\,(nm), $\pm10\%$, AFM   & 82.4 & 185.3 & 250.0 & 325.3  & 355.4  & 458.5  \\ \hline
$t$\,(nm), fitted & 81.5 & 192.7 & 263.3 & 330.6 & 419.3 & 507.3
\end{tabular}
\caption{Thickness $t$ of each MoS$_2$ flake measured with AFM, showing also values from the fitting procedure.} 
\label{tab:t_fit}
\end{table}
\begin{table}[!h]
\centering
\begin{tabular}{l|l|l}
               & \multicolumn{1}{c|}{$\varepsilon_\parallel$} & \multicolumn{1}{c}{$\varepsilon_\perp$} \\ \hline
Ermolaev \emph{et al.}~\cite{Ermolaev:2021} & $16.56$                & $6.43$      \\ \hline
Current work         &          $16.11\pm 0.07$               &      $6.25\pm0.04$          
\end{tabular}
\caption{Permittivity components (Eq.~\ref{eq:MoS2}) of MoS$_2$ at $\lambda_0=1570$\,nm.}
\label{tab:epsilon_fit}
\end{table}

The effective mode indices, calculated using the experimentally determined permittivities, are presented in Figure~\ref{fig:fig3}. These are shown alongside the corresponding curves calculated using permittivity data from Ermolaev~\emph{et~al.}~\cite{Ermolaev:2021} for comparison. Overall, our experimental results (Figure~\ref{fig:fig3} and Table~\ref{tab:epsilon_fit}) show qualitative agreement with previously reported data~\cite{Ermolaev:2021, Munkhbat:2022}. However, at a quantitative level, the results reveal a difference in the effective mode index, most notably for the TE modes. 
One should also note that there is a significant difference between the thicknesses measured with AFM and the fitted ones (Table~\ref{tab:t_fit}), with the discrepancy increasing as the thickness becomes larger. This can partially be attributed to the calibration being done with a calibration grating of 100\,nm high steps, meaning that the thicker flakes may be outside of the calibration region and the scanner is non-linear for those step heights. However, the discrepancy is uncommonly high and non-monotonic, suggesting other causes yet to be found. 

By representing the data in a parametric plot, where the effective mode indices are compared to the one of TM$_0$ mode (for thicknesses where they co-exist),  we eliminate the uncertainties associated with specific values of $t$ (Figure~\ref{fig:fig3}b). This plot clearly shows the difference between our measurements and calculations using the reported optical constants~\cite{Ermolaev:2021}. Moreover, the localization of our measured points close to the fitted lines suggests high accuracy of our method. This demonstrates that the used method is both suitable and capable of determining the permittivity of crystalline TMDC flakes with lateral dimensions as small as tens of microns, which are otherwise challenging to probe using standard ellipsometry techniques due to the limited surface uniformity over larger flake areas and spot size.

\section{Discussion}

This work demonstrates a novel method for determining the optical properties of materials by using transmission-type s-SNOM to measure the real part of its permittivity. By measuring the complex near-field map of mechanically exfoliated MoS$_2$ flakes of varying thicknesses on BK7 glass, we determined the permittivity components of MoS$_2$ with a relative error of $\sim 0.5\%$, while probing only areas of approximately $40\,\mathrm{\upmu m} \times 20\,\mathrm{\upmu m}$.
At the first glance, the estimated error is much smaller than the one expected from the Fourier transform (for our measurements with 40-$\upmu$m-long range, the resolution of Fourier spectrum is $\Delta k/k_0 \approx 0.04$, which is more than $1\%$ of $N_m$ for all studied guided modes). However, the uncertainty in determining the position of the peak (which can be found, for example, by locally fitting with a Lorentzian function) can easily be much smaller than the Fourier resolution, when it is known that it should be a peak of a single mode, and the signal-to-noise level is high enough. Similarly, the frequency of a sine curve can be accurately determined by fitting, even if the interval is less than the sine period itself (i.e., when the resolution of the Fourier spectrum is less than the frequency itself). In fact, each fitting of the guided mode in our measurements resulted in the error of $\Delta\sim 0.001$. However, the variations between measurements with different flake orientations resulted in the error of $\Delta\sim0.01$, according to Equation~\ref{eq:delta}. We attribute this variation to the imperfect synchronization of the bottom parabolic mirror during the scan (see Methods), which can be improved to further increase the accuracy of determining the permittivity. Alternatively, these errors indicate that the probing area can be reduced in the current setup without compromising the accuracy of the method.   

As a proof of concept, we focused on a known low optical loss region of MoS$_2$ and therefore only analyzed the real part of the effective mode indices, which is associated with the real part of the permittivity.
By analyzing the decay in amplitude of $\left|E_\mathrm{nf}\right|$ along the propagation direction, it is possible to determine the imaginary part as well (when the divergence of the guided Gaussian beam is taken into account).
However, in frequency regions where the material exhibits high optical losses, the field will decay rapidly, limiting the extent of the detectable region and thus the precision of the results. In particular, the dispersive coupling to other interactions, such as excitons seen in Figure~\ref{fig:fig2}a, would also influence the decay. 
For sufficiently thick flakes, the near-field signal also decreases in overall strength for some modes, as most of the field becomes confined within the flake itself, making near-field measurements more challenging.

Our results show a slight difference in the effective mode indices of the TE and TM modes compared to the expected values based on literature permittivity values~\cite{Ermolaev:2021,Munkhbat:2022}, indicating a discrepancy in the in-plane ($\varepsilon_\parallel$) and out-of-plane ($\varepsilon_\perp$) permittivity components. The discrepancy in $\varepsilon_\perp$, found to be $\sim3$\% from the ellipsometry data reported by Ermolaev \emph{et al.}~\cite{Ermolaev:2021}, can be ascribed to s-SNOM's high sensitivity to out-of-plane polarized fields. In contrast, ellipsometry of high-index materials suffers from small refracted angles, resulting in small out-of-plane $E$-field components and correspondingly low sensitivity to $\varepsilon_\perp$. Surprisingly, we have also found a difference of $\sim3$\% in $\varepsilon_\parallel$, where ellipsometry is supposed to provide accurate measurements.

In the proposed method, measurements are restricted to a single wavelength at a time, which limits its applicability for broadband determination of optical properties. To overcome this limitation, one could use a broadband source and another detection scheme, known as nano Fourier-transform infrared spectroscopy (nano-FTIR), where one records a full interferogram at each point and consequently applies the Fourier transform to transform it to a spectrum~\cite{Kaltenecker:2021}. However, this would significantly increase the acquisition time for a similarly sized scan area; therefore, there may consequently be a need to decrease the number of lateral sampling points.

While we have illustrated the principle with measurements on MoS$_2$, we emphasize its applicability to the broader class of vdW materials, including topological insulators~\cite{Menabde:2024,Yuan:2017,Dubrovkin2017,Venuthurumilli:2019}, and even non-vdW materials~\cite{Balan:2022}. Additionally, one can modify guided modes by using a different substrate, which might improve the sensitivity. For example, by having a metallic mirror as a substrate, vdW flakes will support plasmon polaritons~\cite{Casses:2024,Iyer:2022} and image polaritons at longer wavelength~\cite{Menabde:2022c,Menabde:2022a,Menabde:2022b}.

In conclusion, the proposed method for determining the permittivity of materials using transmission-type s-SNOM serves as an addition to the well-studied ellipsometry method, specifically, for accurate optical characterization at the microscale.

\section{Methods}

\textbf{Substrate cleaning procedure.}
The SCHOTT N-BK7\textsuperscript{\textregistered} glass substrates are ultrasonicated in acetone followed by isopropyl alcohol (IPA) for 5 minutes each. Acetone can dissolve non-polar and polar compounds (like oil and organic compounds), however acetone leaves residues. IPA removes the remaining acetone and can also dissolve non-polar compounds. The ultrasonication loosens particles and residues adhering to the surface. As a last step the substrates are rinsed with deionized water and dried with pressurized nitrogen (N$_2$).

\textbf{Mechanical Exfoliation procedure.}
To mechanically exfoliate MoS$_2$, residual free wafer-tape (Nitto Denko Corporation) is used. 
The 'mother' crystal is placed in contact with the tape, and when it is pulled away, a significant amount of material is transferred to the tape. By repeatedly sticking and unsticking the tape to itself in the area with the material, the crystals gradually become thinner.
Afterwards, a polydimethylsiloxane (PDMS) stamp (Gel-Pak 8) is brought into contact with a selected area of the tape and then peeled off, leaving flakes on the PDMS.
The PDMS is then placed on a glass slide with the flake facing away from the glass surface. Using a manipulation stage, the glass slide and PDMS are slowly lowered toward a BK7 glass wafer chip mounted on a vacuum chuck at $60^\circ$C, where they eventually make contact (the procedure is monitored through an optical microscope).
Lastly, the PDMS stamp is slowly lifted up, leaving MoS$_2$ flakes on the glass chip.
  
\textbf{Near-field setup.}
The near-field measurements were performed using a customized commercially available transmission type s-SNOM (NeaSpec, Attocube), and a sketch of the setup is given in the SI section S3. The setup employs pseudo-heterodyne demodulation to simultaneously acquire amplitude and relative phase information from the near-field signal. A continuous-wave near-infrared laser beam ($\lambda_0=1570\,\mathrm{nm}$) is split into two paths. One path is the reference arm, where the light is modulated by an oscillating mirror ($f\approx300$\,Hz). In the other path the laser beam is focused ($\sim 3\,\mathrm{\upmu m}$ spot size) onto the edge of a flake by a parabolic mirror (PM) below the sample. A near-field probe (Pt-coated ARROW-NCPt, NanoWorld) scatters the near field into free space, transforming the bound evanescent waves into freely propagating waves. 

The scattered near-field signal is collected by another PM above the sample and is further recombined with reference beam so their interference can be detected. The detected signal is then subsequently demodulated (pseudo-heterodyne detection) at higher harmonics of the probes oscillation frequency ($\eta\Omega$, with $\eta=3$ and $4$) to suppress the background (any light scattered from the tip or the sample, but not related to the probed near field). 

When scanning, the sample is moved and to maintain the excitation beam spot at the flake edge, the bottom PM is moved synchronously with the sample. However, the stage for the bottom PM is not as precise as the sample stage, which mainly results in a small artificial phase ‘wobbling’ of the excitation light. This leads to spectral leakage in the Fourier domain, which can be seen as small sidebands around guided modes (see, for example, TM$_0$ mode in Figure~\ref{fig:fig1}e). This, in turn, can result in the incorrect determination of $N_m$ for closely spaced modes, when their spectral leakage will overlap (for example, TM$_1$ and TE$_2$ for the 460-nm-thick flake). To correct the phase ‘wobbling’ and determine $N_m$ without artificially lowering the estimated errors, we use the following two-step procedure:

\emph{i)} First, we select the mode with the most prominent peak in the Fourier spectrum, which does not overlap with others, and filter it using a square window function with a width of $0.5 k_0$. Then it is inversely transformed back to real space, converted to 1D by performing integration along the $y$-direction, followed by a linear fit of the unwrapped phase. This process provides the residual phase, which is then subtracted from the raw data.

\emph{ii)} In the second iteration, the corrected complex near-field data is Fourier transformed again and filtered for each mode using a smaller square window function with a width of $0.15 k_0$, followed by the same procedures to provide fitted $N_m$. Importantly, to avoid artificial lowering of the uncertainty, we estimate the squared error of $N_m$ for each mode as the sum of the squared error in the second step and the squared error for the reference mode in the first step.

\bibliographystyle{apsrev4-1}
\bibliography{references}

\section{Acknowledgments} %
The Center for Polariton-driven Light--Matter Interactions (POLIMA) is sponsored by the Danish National Research Foundation (Project No.~DNRF165).
C.~F. was supported by the Carlsberg Foundation as an Internationalisation Fellow (Grant No.~CF21-0216), while also acknowledging the VILLUM Foundation (Grant No.~58634).
Authors acknowledge fruitful discussions with S.~I. Bozhevolnyi and G.~A. Ermolaev. 

\section{Author Contributions Statement}

M.~N., V.~A.~Z., and N.~A.~M. conceived the idea. M.~N., T.~Y., and C.~F. contributed to the sample fabrication and materials characterization. Optical measurement were performed by M.~N. and V.~A.~Z., while S.~R. developed the fitting procedure for the dispersion relations.
All authors contributed to analyzing the data and writing the manuscript. 
All authors have accepted responsibility for the entire content of this manuscript and approved its submission.

\clearpage

\widetext
\setcounter{page}{1}
\setcounter{equation}{0}
\setcounter{figure}{0}
\setcounter{table}{0}
\setcounter{section}{0}
\renewcommand{\thepage}{s\arabic{page}}
\renewcommand{\theequation}{S\arabic{equation}}
\renewcommand{\thetable}{S\arabic{table}}
\renewcommand{\thefigure}{S\arabic{figure}}
\noindent\textbf{\large Supplementary Information}
\section{S1. Material permittivities}

\label{sec:supp-permittivity}
The relative permittivity of the BK7 glass substrates are given by the Sellmeier dispersion formula 
\begin{equation}
    \varepsilon_s(\omega)=1+\sum_i\frac{B_{s,i}\lambda^2}{\lambda^2-C_{s,i}},\quad \lambda=2\pi c/\omega,
\end{equation}
where $B_{s,i}$ and $C_{s,i}$ are experimentally determined Sellmeier coefficients. The coefficients for BK7 glass are presented in Table ~\ref{tab:supp-Sellmeier}.

\begin{table}[!h]
\centering
\begin{tabular}{c|c|c}
B$_{s,1}$          & B$_{s,2}$           & B$_{s,3}$           \\ \hline
1.039612120 & 0.231792344  & 1.010469450  \\ \hline
C$_{s,1}$ $(\upmu\mathrm{m}^2)$          & C$_{s,2}$ $(\upmu\mathrm{m}^2)$          & C$_{s,3}$ $(\upmu\mathrm{m}^2)$           \\ \hline
0.006000699 & 0.0200179144 & 103.56065300
\end{tabular}%
\caption{Sellmeier coefficients for BK7 glass~\cite{SCHOTT}.}
\label{tab:supp-Sellmeier}
\end{table}

The relative permittivity of MoS$_2$ used to find the modes in the dispersion diagram (Figure~2 in the main text) is from Ermolaev et. al.~\cite{Ermolaev:2021} where they experimentally determined it using imaging ellipsometry and the Tauc--Lorentz oscillator model. The in-plane relative permittivity is determined by
\begin{equation}
    \varepsilon_{\parallel}(E)=\varepsilon_\infty+\varepsilon_{\mathrm{uv}}(E)+\sum_i\varepsilon_{\mathrm{TL},i}(E),
\end{equation}
where $E$ is the photon energy ($E=\hbar\omega$), $\varepsilon_{\infty}$ is the relative permittivity at infinite photon energy, $\varepsilon_{\mathrm{TL},i}(E)$ is the complex Tauc--Lorentz oscillator function, and $\varepsilon_{\mathrm{uv}}$ is the added ultra-violet (UV) pole:
\begin{equation}
    \varepsilon_{\mathrm{uv}}(E)=\frac{A_{\mathrm{uv}}}{E_{\mathrm{uv}}^2-E^2},
\end{equation}
with the amplitude $A_\mathrm{uv}=228\,\mathrm{eV}^2$ and energy position $E_\mathrm{uv}=15\,\mathrm{eV}$, accounting for strong absorption in the UV.

The Tauc--Lorentz oscillator model for the imaginary part of the in-plane permittivity is related to its real part through the Kramers--Kronig transform
\begin{align}
\mathrm{Im}\left[\varepsilon_{\mathrm{TL},i}(E)\right]=
    \begin{cases}
        \frac{A_i E_{0,i} \Gamma_i (E-E_{g,i})^2}{(E^2-E_{0,i}^2)^2+\Gamma_i^2E^2}\cdot\frac{1}{E},& E>E_{g,i}\\
        0,              & E\leq E_{g,i}
    \end{cases},\\
\mathrm{Re}\left[\varepsilon_{\mathrm{TL},i}(E)\right]=\frac{2}{\pi}\int_{E_g}^\infty \frac{\xi\,\mathrm{Im}\left[\varepsilon_{\mathrm{tl},i}(E)\right]}{\xi^2-E^2}\mathrm{d}\xi,
\end{align}
where $A$ is the oscillator (exciton) strength (amplitude) of the peak, $\Gamma$ is the broadening term of the peak, $E_g$ is the material's optical bandgap energy, and $E_0$ is the peak central energy. The values used are collected in Table~\ref{tab:supp-TL_param}.

\begin{table}[!h]
\begin{tabular}{c|c|c|c|c}
$i$  & $A$ (eV) &  $E_0$ (eV) & $\Gamma$ (eV) & $E_g$ (eV)           \\ \hline
1 & 308    & 1.852   & 0.067   & 1.761              \\ \hline
2 & 135    & 2.006   & 0.148   & 1.82               \\ \hline
3 & 19.3   & 2.662   & 0.380   & 1.24               \\ \hline
4 & 69     & 2.99    & 1.348   & 1.31              
\end{tabular}
    \caption{Tauc--Lorentz oscillator model parameters provided by Ermolaev et al.~\cite{Ermolaev:2021}.}
    \label{tab:supp-TL_param}
\end{table}

\begin{figure}[!h]
    \centering
    \includegraphics[width=0.5\linewidth]{  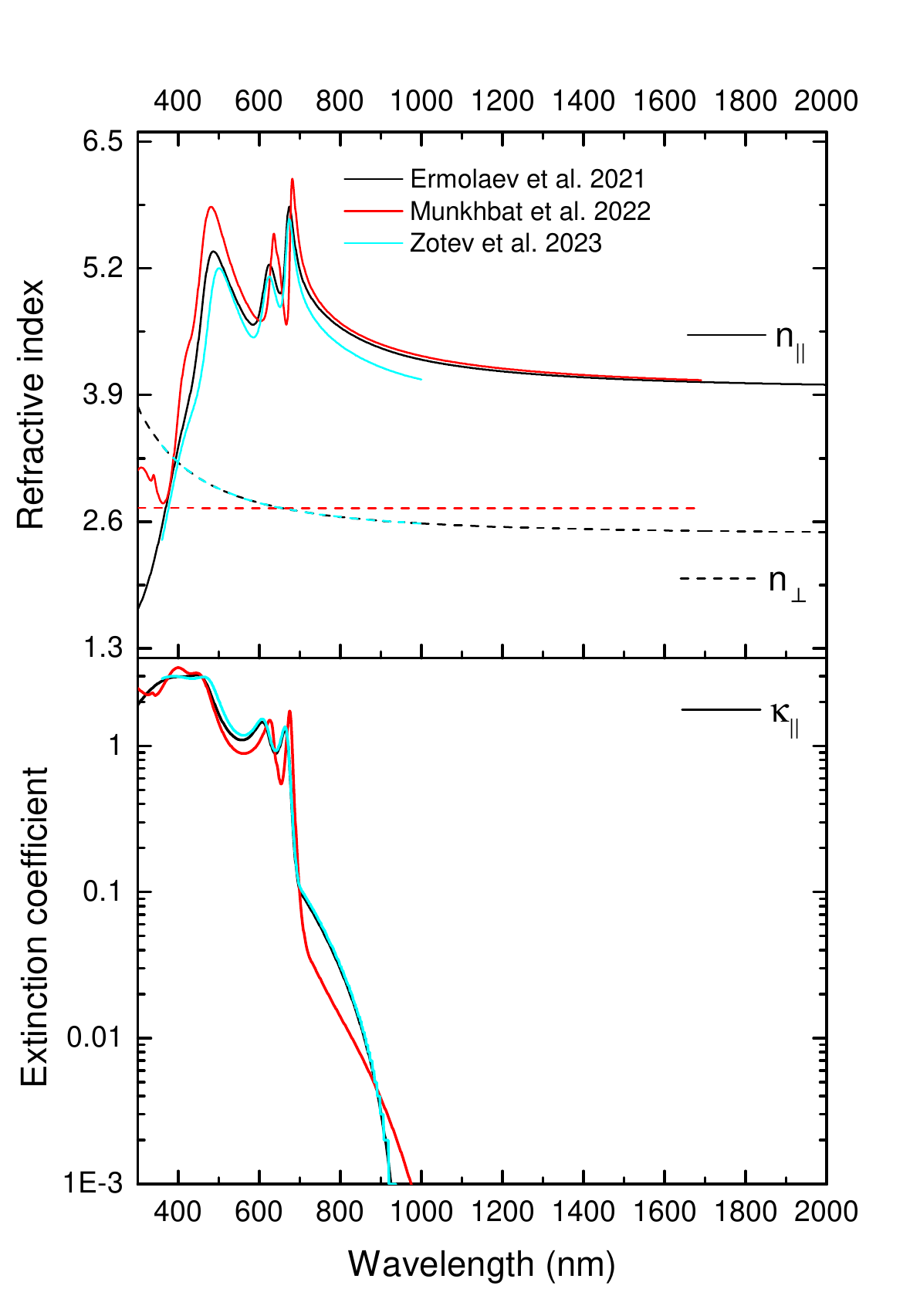}
    \caption{Refractive index ($n$) and extinction coefficient ($\kappa$) of MoS$_2$ from Ermolaev et al.~\cite{Ermolaev:2021}, Munkhbat et al.~\cite{Munkhbat:2022}, and Zotev et al.~\cite{Zotev:2023}. There is a notable difference between measured refractive indices even in the low-loss region ($\lambda>700$ nm). Imaging ellipsometry was used in \cite{Ermolaev:2021} and \cite{Zotev:2023}, while a conventional ellipsometry setup was used in \cite{Munkhbat:2022}.}
    \label{fig:supp-nk_MoS2}
\end{figure}

For the out-of-plane component we use Cauchy's equation with two terms
\begin{align}
    \mathrm{Re}\left[\varepsilon_\perp (\lambda)\right]=\left(A_c+\frac{B_c}{\lambda^2}\right)^2,\\
    \mathrm{Im}\left[\varepsilon_\perp (\lambda)\right]=0,
\end{align}
where $A_c=2.463$ is the refractive index at infinite wavelength and the coefficient $B_c=119\cdot10^3 (\mathrm{nm}^2)$ accounts for the first-order wavelength dependency.
The resulting permittivities in the form of refractive index ($n$) and extinction coefficient ($\kappa$), $\varepsilon=\left(n+i\kappa\right)^2,$ are shown in Figure~\ref{fig:supp-nk_MoS2}, together with two other measurements from the literature.

\clearpage

\section{S2. SNOM probing sensitivity to different modes}
\label{sec:supp-Mode_influence}
\begin{figure}[!h]
    \centering
    \includegraphics[width=\linewidth]{  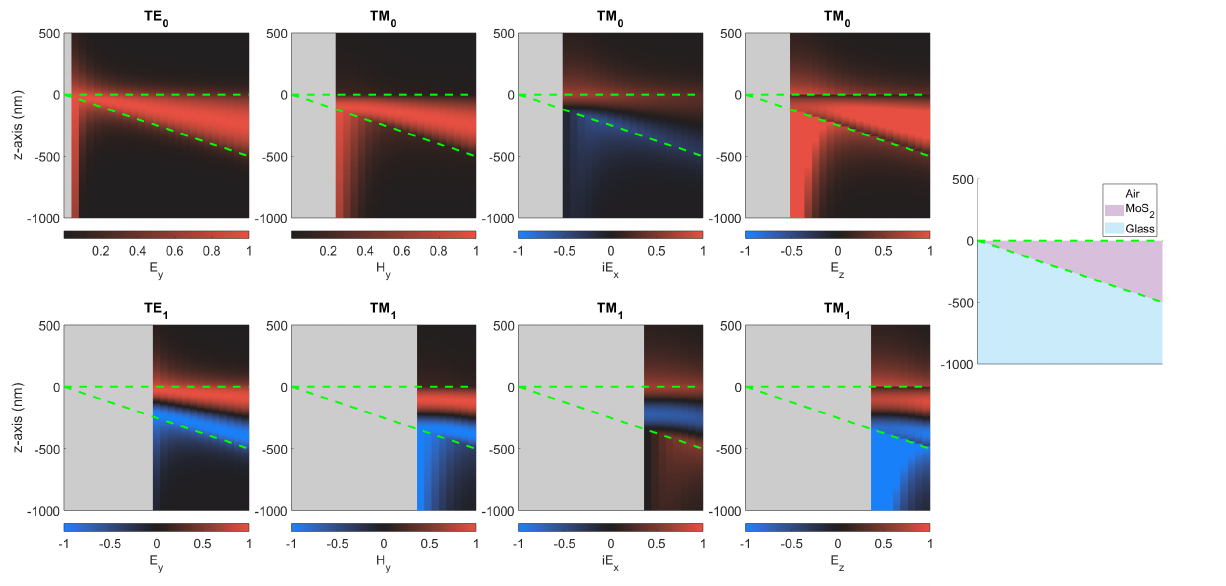}
    \caption{Field distributions for TE$_m$ ($m$=0,1) and TM$_m$ ($m$=0,1) modes for varying thickness of MoS$_2$. The gray area is the region of leaky modes. The field of TE modes are normalized by the maximum of $E_y$ for each thickness. Similarly, TM modes are normalized by the maximum of $H_y$ for each thickness. Thus, one can directly compare $E_x$ and $E_z$ fields in strength and verify the dominance of the normal $E_z$ component for TM modes.}
    \label{fig:supp-Mode_dist}
\end{figure}

The field distribution of the field components for the first two orders of the TE and TM modes for each thickness is depicted in Figure~\ref{fig:supp-Mode_dist} as a color gradient.

The mode distribution significantly influences the detectability of modes when using s-SNOM. The scattered signal in s-SNOM is dependent on the orientation of the electric field components relative to the tip, with the strongest signal occurring when the components are parallel. This is because the interaction between the tip and the electric field is maximized when they are aligned. For TE modes, the electric field has a single in-plane component ($E_y$), which substantially reduces the scattering and collection efficiency. Conversely, for TM modes, the electric field components include both in-plane ($E_x$) and out-of-plane ($E_z$) elements, resulting in relatively higher scattering efficiency.

The detectability of modes is also influenced by the mode order and waveguide thickness. Higher-order modes, with more complex field distributions, typically exhibit weaker confinement and lower detectability compared to fundamental modes. Thinner waveguides enhance the interaction between the extended electromagnetic field and the s-SNOM tip, improving detectability. However, as the waveguide thickness increases, a larger portion of the electromagnetic field is confined within the waveguide, reducing the field's interaction with the tip and thus decreasing detectability. These effects are crucial for accurately interpreting results.

\clearpage

\section{S3. MoS\textsc{2} flake characterization}
%\label{sec:supp-characterization}

\subsection{Imaging of flakes}

\begin{figure}[!h]
    \centering
    \includegraphics[width=\linewidth]{  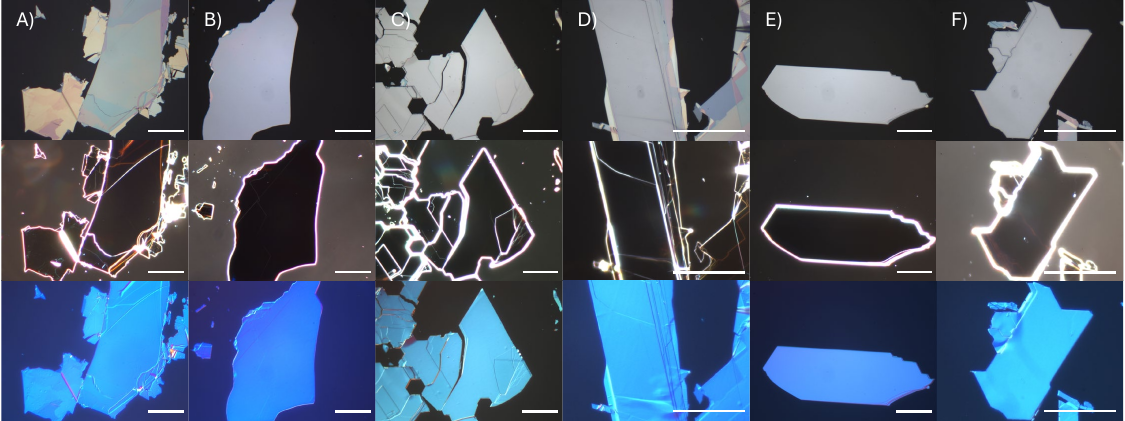}
    \caption{Bright-, dark-field, and DIC images of the flakes measured. From \textbf{A)} to \textbf{F)} the thicknesses are as measured by AFM given in Table \ref{tab:supp-AFM}. The scale bars are $50\,\mu\mathrm{m}$.}
    \label{fig:supp-all_flakes}
\end{figure}

The mechanically exfoliated MoS$_2$ flakes on BK7 glass are characterized by conventional optical microscopy using a optical microscope (Axiotech, Zeiss) which allows for bright-field (BF), dark-field (DF) and differential interference contrast (DIC) imaging. The images of the measured flakes are presented in Figure~\ref{fig:supp-all_flakes}. 

\subsection{AFM measurements of flake thickness}
The thickness of each flake is determined using AFM by scanning over the flake edge (the edge is parallel to the slow scan $y$-axis) in several slightly different locations. The raw data from each scan is corrected for tilt and offset (line shift) as the following:
\begin{equation}
    z(x,y) = z_\mathrm{raw}(x,y) - ax - b(y),
\end{equation}
where slope $a$ and offset $b(y)$ are found by minimizing $\sum z^2$ for points corresponding to the substrate. The resulting thicknesses for each flake are collected in Table~\ref{tab:supp-AFM}. 
\begin{table}[!h]
    \begin{tabular}{l|l|l|l|l|l|l}
    Flake          & A     & B      & C   & D     & E     & F   \\ \hline
    Thickness (nm) & 82.4 & 185.3 & 250 & 325.3 & 355.4 & 458.5
    \end{tabular}
    \caption{AFM measured thicknesses.}
    \label{tab:supp-AFM}
\end{table}

\subsection{Raman spectroscopy}
\begin{figure}[!h]
    \centering
    \includegraphics[width=0.5\linewidth]{  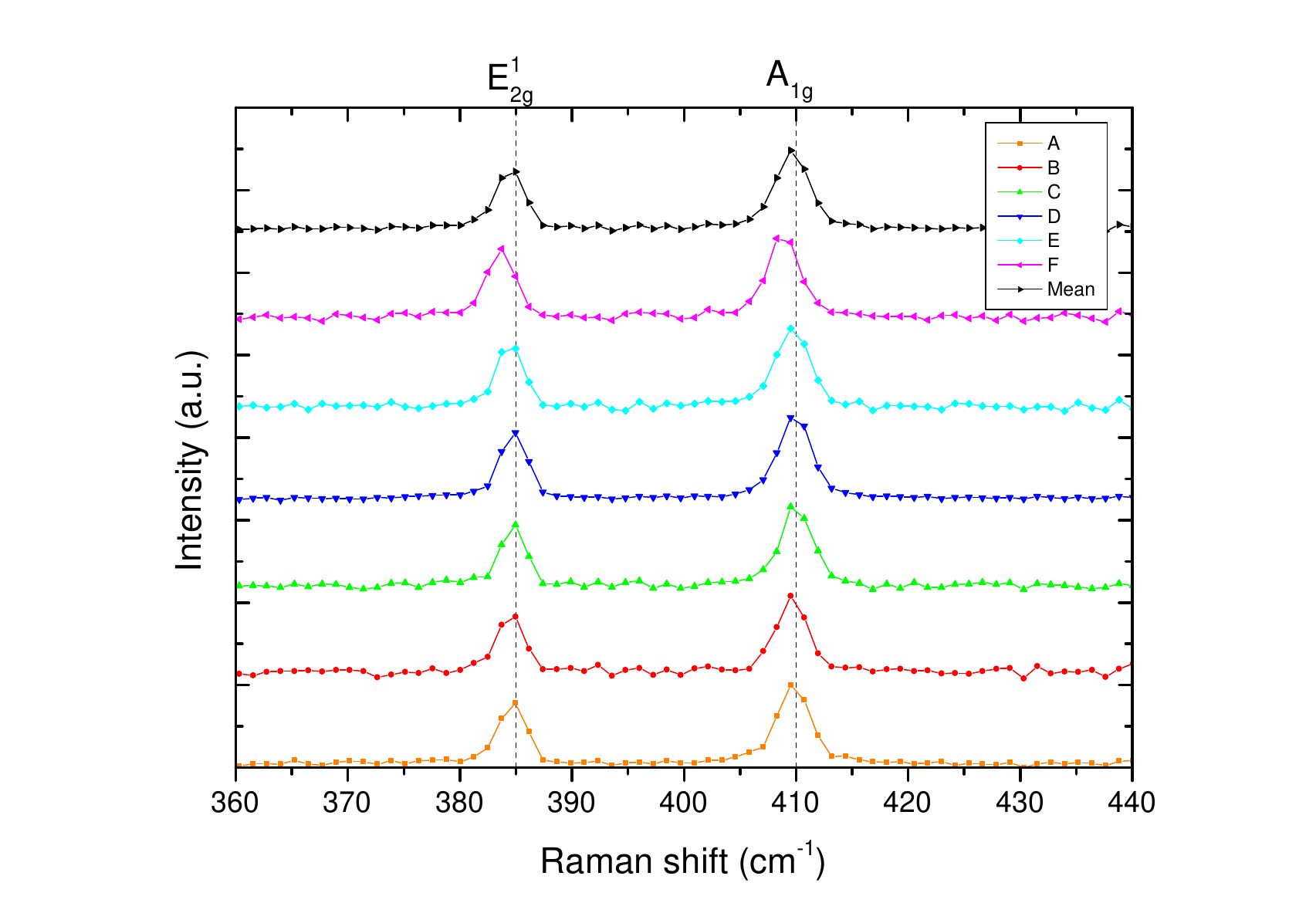}
    \caption{Raman spectra of the individual flakes on BK7-glass and their MoS$_2$ Raman-modes using a laser with a (non-excitation) wavelength of $532\,\mathrm{nm}$. Literature values (vertical dashed lines) for Raman-modes of MoS$_2$ are taken from Carvalho et al.~\cite{Carvalho:2015}.}
    \label{fig:supp-raman_spec}
\end{figure}

To verify that the flakes are made of MoS$_2$ (HQ-graphene, 2H-MoS$_2$ natural crystal) we use Raman spectroscopy to compare the peaks in the Raman shift to values given in the literature. Figure~\ref{fig:supp-raman_spec} shows the region where the in-plane $E_{2g}^1$ and out-of-plane $A_{1g}$ Raman modes are active. 
  
Bulk MoS$_2$ exists in different crystalline phases (the stacking of the individual layers is different): 1T (octahedral symmetry), 2H (hexagonal symmetry), and 3R (rhombohedral symmetry). The Raman spectrum permits us to distinguish between these since the 1T-phase has Raman modes that are different from the two other phases. The 2H and 3R phases have peaks at $\sim E_{2g}^1=385\,\mathrm{cm}^{-1}$ and $\sim{A_{1g}}=410\,\mathrm{cm}^{-1}$ for the in-plane and out-of-plane vibrational modes respectively (with a difference between the two peaks of $\sim25\,\mathrm{cm}^{-1}$)~\cite{Carvalho:2015}. For the 1T-phase the in-plane $E_{2g}$ mode is absent and out-of-plane $\sim{A_{1g}}=405\,\mathrm{cm}^{-1}$~\cite{Fang:2018}. 
To distinguish the 3R phase from the 2H phase one can look at the second harmonic generation that is present for bulk 3R-MoS$_2$ because of it being non-centrosymmetric (lacks inversion symmetry).

MoS$_2$ show different optical properties depending on the different crystalline phases and its thickness when it approaches a few layers (each layer is $\sim0.65\,\mathrm{nm}$ in thickness). However, we only considered bulk flakes and can thus neglect the thickness dependent intrinsic optical material properties.

\subsection{Thickness measurements by reflection spectroscopy}
Another way to characterize the MoS$_2$ flakes is with reflection spectroscopy, where it's possible to extract the thickness of the flakes assuming the permittivity is known (or vice versa). To this end we need the Fresnel equations for the case of a uniaxial crystal. A sketch of the configuration is shown in Figure~\ref{fig:supp-fresnel}.

\begin{figure}[!h]
    \centering
    \begin{tikzpicture}
        \filldraw[thick,color=black!100, fill=magenta!10] (0,0) rectangle (4,1);%MoS2

        %Glass
        \filldraw[color=white,shading = axis,rectangle,left color=blue!15,shading angle=0]   (-0.5,0) rectangle (4.5,-1);
        \draw[very thick] (-0.5,0)--(4.5,0);

        % arrows
        \draw[-{latex}] (0.5,3)--(1.5,1)--(2.5,3);
        \draw (1.5,1)--(1.75,0)--(2,1)--(2.25,0)--(2.5,1)--(2.75,0);
        \draw[-{latex}] (1.5,1)--(2.5,3);
        \draw[-{latex}] (1.75,0)--(2.2,-1);
        \draw[-{latex}] (2.25,0)--(2.7,-1);
        \draw[densely dotted] (2.25+0.5,0)--(2.48+0.5,-0.5);
        \draw[-{latex}] (2,1)--(3,3);
        \draw[-{latex}] (2.5,1)--(3.5,3);
        \draw[densely dotted] (2.75,0)--(2.75+0.14,0.5);
        \draw[<->] (4.2,0)--(4.2,1) node [midway,right] {$t_{\mathrm{MoS}_2}$};
        \draw[dashed] (1.5,0)--(1.5,2.5);
        \draw[dashed] (1.75,0)--(1.75,-1);
        \draw[thick] (1.5,1.5) .. controls (0.875+0.5,1.525) .. (1.25,1.5) node [midway, above] {$\theta_i$};
        \draw[thick] (1.5,0.5) .. controls (1.57,0.475) .. (1.62,0.5) node [midway, left] {$\theta_{2}$}; 
        \draw[thick] (1.75,-0.5) .. controls (1.88,-0.525) .. (1.98,-0.5) node [midway, left] {$\theta_{3}$};         
        \draw (-0.25,1.75) node {$n_1$};
        \draw (0.35,0.5) node {$n_2$};
        \draw (-0.25,-0.75) node {$n_3$};
        \draw (2.5,3) node [above] {$\underline{r}$};
        \draw (3,3) node [above] {$\underline{r}_0$};
        \draw (3.5,3) node [above] {$\underline{r}_1$};
        \draw (4,3) node [above] {$\ldots$};
        \draw[->] (0.5,3)--(0.75,3.125) node[at start] {$\odot$};
        \draw[] (0.75,3.125) node[right] {$E_p$};
        \draw[] (0.5,3) node[left] {$E_s$};
    \end{tikzpicture}
    \caption{A plane wave incident on a thin dielectric layer of thickness $t$ on a semi-infinite thick substrate.}
    \label{fig:supp-fresnel}
\end{figure}

Generally, the refractive indices, $n_i(\omega)$, vary with frequency. In a uniaxial crystal, the refractive index encountered by a \textit{p}-polarized plane wave propagating within the crystal is influenced by the refracted angle $\theta_2$, which in turn depends on the incidence angle $\theta_i$ and the refractive indices of the first layer and both components of the second layer. In contrast, for an \textit{s}-polarized plane wave, the electric field interacts solely with the in-plane components of the layers. These factors, along with the Fresnel equations for the complex amplitude reflection coefficients for \textit{s}- and \textit{p}-polarized light at the two interfaces, are expressed as follows:

\begin{align*}
     &\begin{array}{c@{\hspace{2cm}}c}
        \text{\textit{s}-polarization equations} \\
    \end{array} & &\begin{array}{c@{\hspace{2cm}}c}
        \text{\textit{p}-polarization equations} \\
    \end{array} \\
    n_2&=n_\parallel, & n_2&=\left({\frac{\cos^2{(\theta_2)}}{n_\parallel^2}+\frac{\sin^2{(\theta_2)}}{n_\perp^2}}\right)^{-\frac{1}{2}},\\
    \sin{(\theta_2)}&=\frac{n_1}{n_\parallel}\sin{(\theta_i)}, & \sin{(\theta_2)}&=\left({1+\frac{n_\parallel^2}{n_1^2\sin^2{(\theta_i)}}-\frac{n_\parallel^2}{n_\perp^2}}\right)^{-\frac{1}{2}}, \\
    r_{s,12}&=\frac{n_1\cos{(\theta_i)}-n_2\cos{(\theta_{2})}}{n_1\cos{(\theta_i)}+n_2\cos{(\theta_{2})}},          &  r_{p,12}&=\frac{n_1\cos(\theta_{2})-n_2\cos(\theta_i)}{n_1\cos(\theta_{2})+n_2\cos(\theta_i)},\\
    r_{s,23}&=\frac{n_2\cos{(\theta_{2})}-n_3\cos{(\theta_{3})}}{n_2\cos{(\theta_{2}}+n_3\cos{(\theta_{3})}},       &  r_{p,23}&=\frac{n_2\cos(\theta_{3})-n_3\cos(\theta_{2})}{n_2\cos(\theta_{3})+n_3\cos(\theta_{2})},\\
    \label{eq:supp-r_coef}  
\end{align*}
where the subscript numbers indicate the mediums of the interfaces (e.g. $r_{s,12}$ is the reflection from the interface between medium 1 and 2 where the light comes from medium 1). 
The light gets refracted and reflected from each interface in a series of events and gives the complex reflection coefficients: 
\begin{align}
    \underline{r}&=r_{12},\nonumber\\
    \underline{r}_0&=t_{12}r_{23}t_{21}\mathrm{e}^{2i\beta}, \nonumber\\
    \underline{r}_1&=t_{12}r_{23}t_{21}\mathrm{e}^{2i\beta}r_{21}r_{23}\mathrm{e}^{2i\beta}, \nonumber\\
    &\quad\vdots\nonumber\\
    \underline{r}_m&=t_{12}r_{23}t_{21}\mathrm{e}^{2i\beta}\left(r_{21}r_{23}\mathrm{e}^{2i\beta}\right)^{m},
\end{align}
where the total sum ($\underline{r}_{tot}=\underline{r}+\underline{r}_0+\underline{r}_1+\ldots$) is a geometric series that converges to
\begin{equation}
    \underline{r}_{tot}=\frac{r_{12}+r_{23}\mathrm{e}^{2i\beta}}{1+r_{12}r_{23}\mathrm{e}^{2i\beta}},\quad \beta=\frac{2\pi n_2t_{\mathrm{MoS}_2}\cos{(\theta_{2})}}{\lambda_0},
    \label{eq:supp-r_tot}
\end{equation}
for $s$- and $p$-polarized light using the appropriate coefficients from above. The reflectance is given as $R=\left|\underline{r}_{tot}\right|^2$ and for unpolarized light one can use, $R=\left| \underline{r}_{s,tot}\right|^2/2+ \left|\underline{r}_{p,tot}\right|^2/2$, instead.
The total reflectivity can also be found by integrating the reflectance over all the angles ($0\rightarrow\pi/2$) as in the following:
\begin{equation}
    R_{NA}=\int_0^{\pi/2} R\left(\theta\right)F(\theta) \,d\theta,
\end{equation}
where $F(\theta)$ is used to take into account the numerical aperture (NA) of the objective. We assume $F(\theta)$ to have a Gaussian distribution of
\begin{equation}
    F(\theta)=\frac{2}{\sqrt{2\pi\theta_i^2}}\exp{\left[-\left(\frac{\theta}{\theta_i}\right)^2\frac{1}{2}\right]},
\end{equation}
where $\theta_i$ corresponds to the collection half-angle of the objective:
\begin{equation}
    \theta_i=\sin^{-1}\left(\frac{\mathrm{NA}}{n_1}\right).
\end{equation}

\begin{figure}[!h]
    \centering
    \includegraphics[width=\linewidth]{  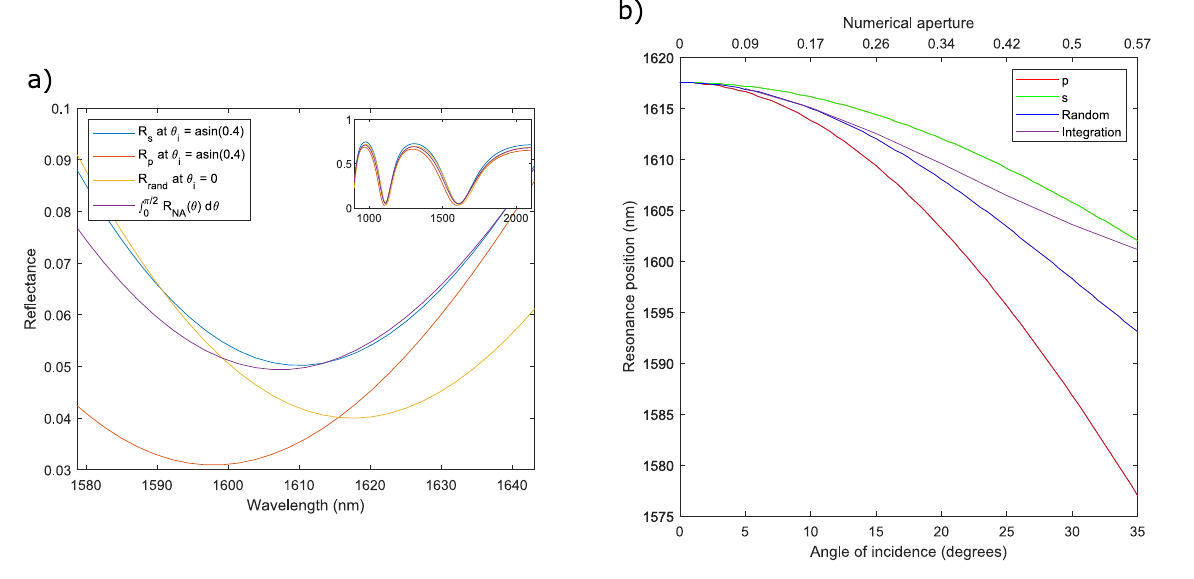}
    \caption{\textbf{a)} Reflectance spectra in a spectral region around a resonance, calculated for normal and tilted incidence for \textit{s}- and \textit{p}-polarized light, compared with integrated un-polarized reflectance. The insert shows the reflectance over the entire wavelength region considered in the measurements. \textbf{b)} The influence of the incident angle (and corresponding NA for the integral form of $R_{NA}$) on the blue shift of the resonance.}
    \label{fig:angle_influence}
\end{figure}

In order to see the influence of the incident angle, we compare reflectance spectra for \textit{s}- and \textit{p}-polarized light at normal incidence and at the angle, corresponding to the collection half-angle of our objective (Figure~\ref{fig:angle_influence}). As expected, inclined incidence results in a blue shift of the resonances, which is smaller for the \textit{s}-polarization (partially because of the corresponding larger refractive index of the flake and, thus, smaller refracted angle $\theta_2$). When we take into account the collection by the objective and integrate the reflectance, then the resultant spectrum features a moderate blue shift. The dependence of the resonance blue shift on the incident angle and corresponding NA for the integral form of $R_{NA}$ is illustrated in Figure~\ref{fig:angle_influence}b.

\begin{figure}[!h]
    \centering
    \includegraphics[width=\linewidth]{  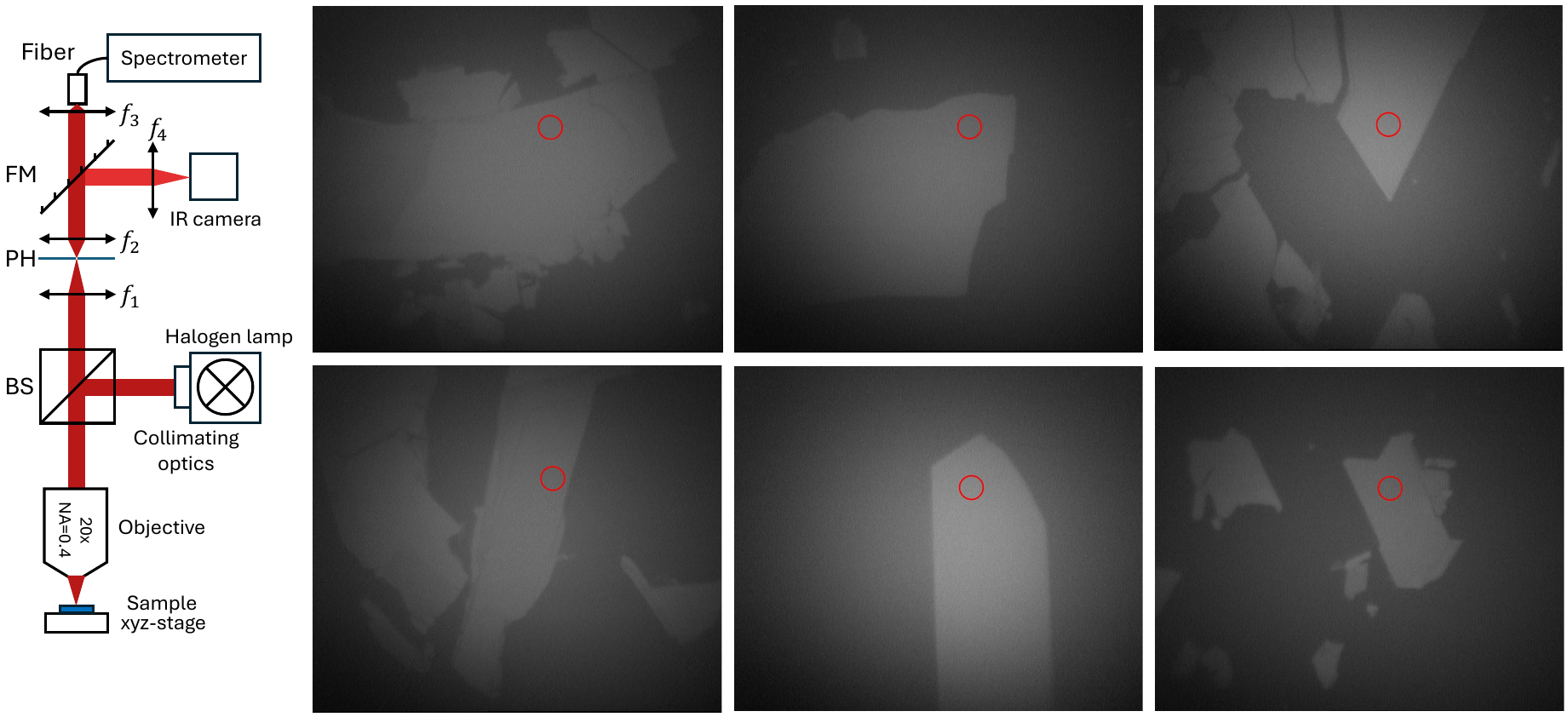}
    \caption{Left, spectroscopy set-up, with the focal length of the lenses $f_1=20$ cm, $f_2=10$ cm, $f_3=3$ cm, and $f_4=20$ cm. The abbreviations are beam-splitter (BM), $300\,\mathrm{\upmu m}$ pinhole (PH), and flip-mount mirror (FM). The pinhole is used to spatially filter the area of the sample, from which the spectrum is collected, and it roughly corresponds to a circle of $15\,\mathrm{\upmu m}$ in diameter (because of 20$\times$ magnification). Right, IR images of flakes A-F with red circles marking the sample area, where the reflection spectra were collected.}
    \label{fig:supp-r_spec_pos}
\end{figure}

In the experiment, we used a halogen lamp as the light source, a near-infra-red spectrometer (NIRQUEST, Ocean Optics, grating NIR2 900-2200 nm, slit $50\,\upmu$m), and a $\mathrm{NA}=0.4$ objective (LMPlan IR 20$\times$/0.40) together with various optical components to obtain the spectra. The setup and the measured area of each flake are shown in Figure~\ref{fig:supp-r_spec_pos}.

\begin{figure}[!h]
    \centering
    \includegraphics[width=1\linewidth]{  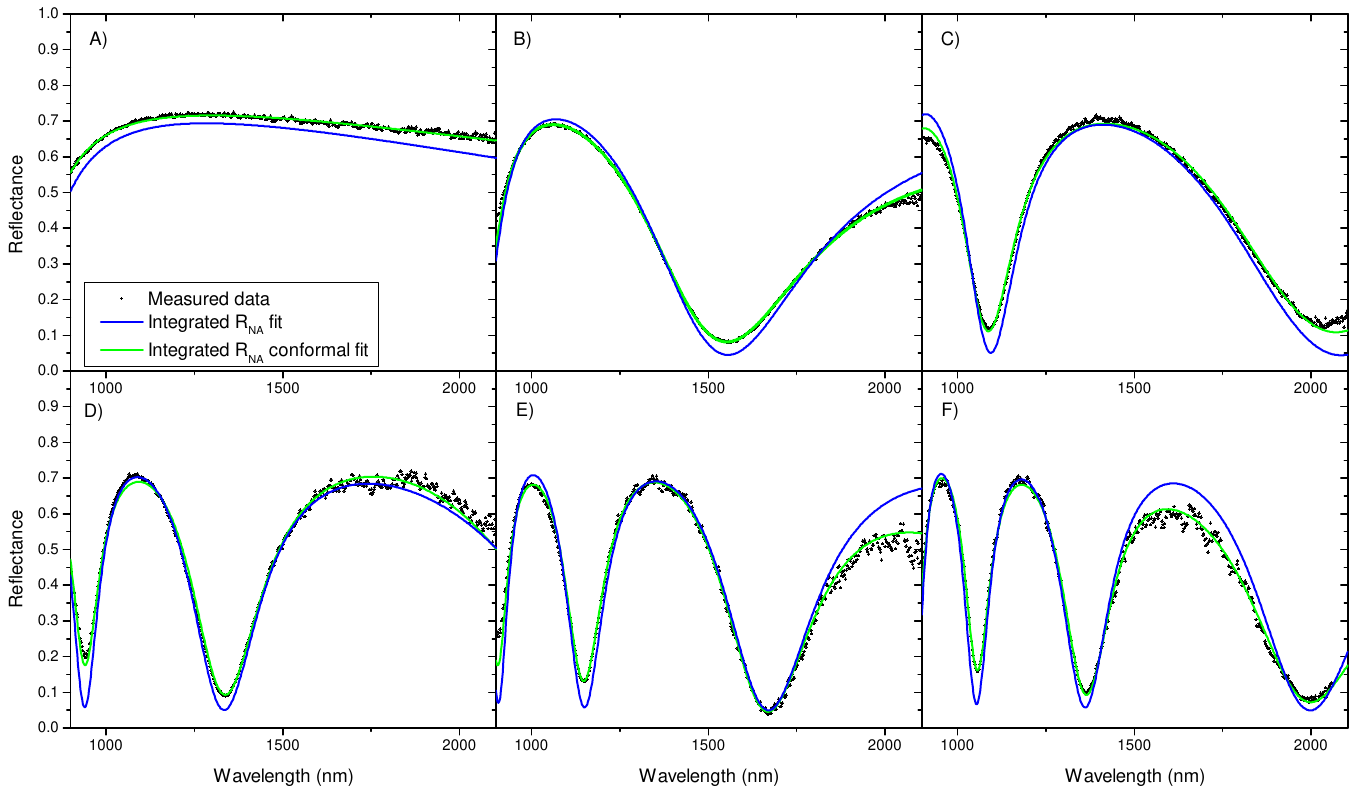}
    \caption{Reflection spectroscopy of the different samples. The assumed refractive index is from Ermolaev et al.~\cite{Ermolaev:2021}. The conformal fits are done with the following fixed parameters: \textbf{A)} $p_{2,3,5,6} = 0$, \textbf{B)} $p_{2,3,6} = 0$, \textbf{C)} $p_{3,6} = 0$, \textbf{D)} $p_{3,6} = 0$, \textbf{E)} $p_3 = 0$, and \textbf{F)} no fixed parameters.}
    \label{fig:supp-r_spectro}
\end{figure}

\begin{figure}[!h]
    \centering
    \includegraphics[width=0.5\linewidth]{  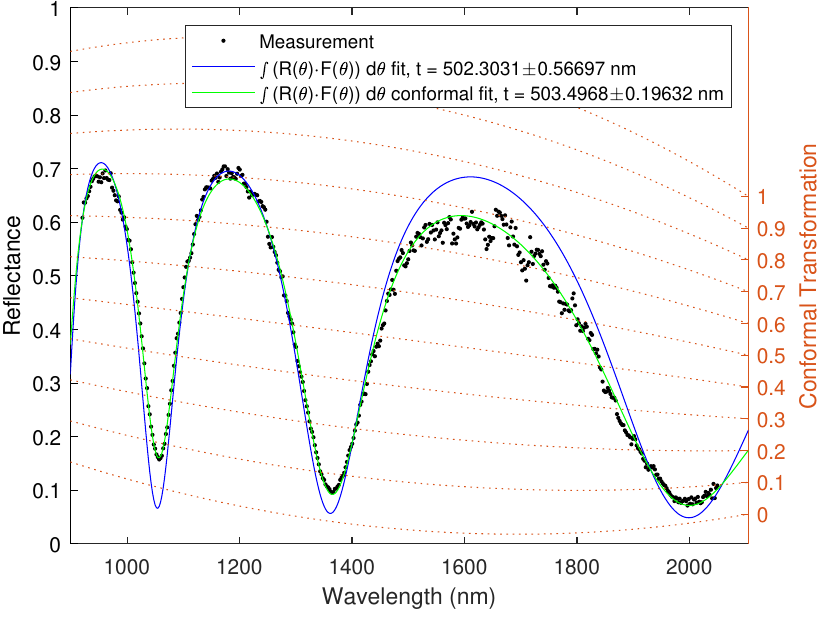}
    \caption{Influence of the conformal transformation. The measured reflectance of flake F (dots), fitted with the integrated reflectivity $R_{NA}$ (solid blue line) and with the conformally transformed $R_{NA}$ (solid green line). The conformal transformation of different reflectance levels are depicted in the right orange scale. As such, the conformal fit does not result in significantly different fitted thickness, but it adjusts the fitted curve to align more closely with the measured points, enhancing the reliability of its results.}
    \label{fig:supp-c_transform}
\end{figure}

The measured reflectance spectra from each flake were normalized by the reflectance from a gold mirror (Thorlabs), preceded by the subtraction of the dark counts background (Figure~\ref{fig:supp-r_spectro}). Usually optical spectra are used to find the position of minima/maxima, because small misalignments, scattering, imperfections, instability of the source spectra, or inconsistent background in the system causes the absolute values to be slightly higher or lower than expected by theory. To overcome this and use all of the spectral data (data points below 920 nm and above 2050 nm are excluded in the fitting due to low signal to noise ratio), we introduce a few extra fitting parameters to conformally transform the  spectra reflectance-axis. Essentially, we would like to transform both the bottom ($R(\lambda) = 0$) and the top ($R(\lambda)=1$) into slowly varying function $f(\lambda)$, which we selected as following:
\begin{equation}
    f(\lambda)=p_i+p_j\sin{(\pi\lambda_{\rm norm}/10+p_k)},
\end{equation}
where $p_{i,j,k}$ are free fitting parameters and 
\begin{equation}
\lambda_{\rm norm} = \frac{\lambda-\lambda_{\rm min}}{\lambda_{\rm max}-\lambda_{\rm min}} - 0.5, 
\end{equation}
is the normalized wavelength. As one can see, this definition ensures slow variation of $f(\lambda)$. This function allows the flexibility in terms of what transformation is needed: a simple offset can be defined by fixing $p_{j}=0$ and $p_{k}=0$, while tilt is achieved by fixing $p_{k}=0$.
Therefore, we transform our theoretical spectra into
\begin{equation}
    \begin{split}
    M(\lambda) &= f_1(\lambda) + R_{NA}f_2(\lambda)\\
    &= p_1+p_2\sin{(\pi\lambda_{\rm norm}/10+p_3)}+R_{NA}(\lambda)\left[1+p_4+p_5\sin{(\pi\lambda_{\rm norm}/10+p_6)}\right],
    \end{split}
\end{equation}
where $p_{1-6}$ are free fitting parameters, which is then used to fit our experimental spectra (Figure~\ref{fig:supp-r_spectro}). The form of the conformal transformation was selected to rescale the reflectivity curve without severely changing the position of its extrema, which can be confirmed in a detailed Figure~\ref{fig:supp-c_transform}.

The results of both the fit with $R_{NA}(\lambda)$ and the conformal fit with $M(\lambda)$ are displayed in Table~\ref{tab:R_spec_fit_values}. As one can see, the fitting results of the near-field measurements coincide much better with the ones of fitting the reflectance spectra, compared to AFM measurements (especially for thick flakes E and F). The introduction of the conformal transform for fitting the reflection spectra has a tiny influence on the fitted thickness, but it reduces the estimated error due to a better match between the measured points and fitted curve. It should be kept in mind, however, that the fitting of the reflection spectra relies on \textit{a priori} knowledge of the flake permittivity. As we found in the near-field study, both the in-plane and out-of-plane permittivity of MoS$_2$ is $\sim3$\% smaller than the permittivity from Ermolaev et al.~\cite{Ermolaev:2021}, used in the reflectance fit. Therefore, if we assume the same difference of $\sim1.5$\% in the refractive index for the whole wavelength range and redo the reflection fit, then the fitted thickness of MoS$_2$ flakes will be $\sim1.5$\% larger than the ones shown in Table~\ref{tab:R_spec_fit_values} (because the position of spectral minima/maxima depends strongly on the optical path length inside the flake, $n_2\,t_{\mathrm{MoS_2}}$), agreeing even better with the results of the near-field fit. This indirectly proves the accuracy of both our near-field and far-field reflectivity fitting methods.
However, one should keep in mind that in both the reflectance fit and the near-field fit the thickness and permittivity of the flakes are linked/coupled via the governing equations, while the AFM measurement depends only on the flake thickness (and on the mechanical properties of the material).

\begin{table}[!h]
\centering
\begin{tabular}{l|l|l|l|l|l|l}
Thickness in nanometer & \multicolumn{1}{c|}{Flake A} & \multicolumn{1}{c|}{Flake B} & \multicolumn{1}{c|}{Flake C} & \multicolumn{1}{c|}{Flake D} & \multicolumn{1}{c|}{Flake E} & \multicolumn{1}{c}{Flake F} \\ \hline
AFM measurements $\pm10\%$ & $82.4$ & $185.3$ & $250.0$ & $325.3$ & $355.4$ & $458.5$   \\ \hline
Reflectance fit           & $81.77\pm0.69$         &$193.26\pm0.21$                        & $262.50\pm0.34$        & $327.73\pm0.35$        & $416.76\pm0.46$        & $502.30\pm0.57$       \\ \hline
Reflectance conformal fit & $82.84\pm0.14$         & $192.85\pm0.05$                       & $260.62\pm0.12$        & $328.02\pm0.19$        & $415.61\pm0.15$        & $503.50\pm0.20$       \\ \hline
Near-field fit & $81.5$ & $192.7$ & $263.3$ & $330.6$ & $419.3$ & $507.3$  
\end{tabular}
\caption{Thickness of each MoS$_2$ flake as experimentally determined from AFM, Reflectance spectroscopy, and SNOM measurements and fits.}
\label{tab:R_spec_fit_values}
\end{table}

\clearpage

\section{S4. Transmission s-SNOM}
\label{sec:supp-sSNOM}
The near-field measurements were performed using a customized commercially available transmission type s-SNOM (NeaSpec, Attocube). The setup (Figure~\ref{fig:supp-sSNOM}) employs pseudo-heterodyne demodulation to simultaneously acquire amplitude and relative phase information from the near-field signal. A continuous-wave near-infrared laser beam ($\lambda_0=1570\,\mathrm{nm}$) is split into two paths. One path is the reference arm, where the light is modulated by an oscillating mirror (RM, $f\approx300$\,Hz). In the other path the laser beam is focused ($\sim 3\,\mathrm{\upmu m}$ spot size) onto the edge of a flake by a parabolic mirror (PM) below the sample. A near-field probe (Pt-coated ARROW-NCPt, NanoWorld) scatters the near field into free space, transforming the bound evanescent waves into freely propagating waves. 

The scattered near-field signal is collected by another PM above the sample and is further recombined with the reference beam so their interference can be detected. The detected signal is then subsequently demodulated (pseudo-heterodyne detection) at higher harmonics of the probe's oscillation frequency ($\eta\Omega$, with $\eta=3$ and $4$) to suppress the background (any light scattered from the tip or the sample, but not related to the probed near field). 

When scanning, the sample is moved; therefore, to maintain the excitation beam spot at the flake edge, the bottom PM is moved synchronously with the sample.
 
\begin{figure}[!h]
    \centering
    \begin{tikzpicture}
        \fill[blue!30] (-3,1) rectangle (1.5,2); % Glass
        \fill[violet!75] (-2,2) rectangle (0.5,2.05); % TMDC
   
        % Source
        \draw[line width=5pt,color=red] (0.5,0)--(2,0);
        \draw[{latex}-,line width=5pt,color=red] (1.5,0)--(8,0) node[right,above] {$\lambda_0=1570\,\mathrm{nm}$};
        \draw[-{latex},line width=5pt,color=red] (3,0)--(3,1.5);
        \draw[-{latex},line width=5pt,color=red] (3,1)--(3,2)--(4,2)--(4,3)--(3,3)--(3,4.5);
        \draw[-,line width=5pt,color=red] (3,3) -- (3,5); % midway from RM to top BS
        \draw[-{latex},line width=5pt,color=red] (3.05,5) -- (6,5); % Top BS to detector
        \filldraw[color=red] (0.6,0) -- (0.4,0) -- (0.5,2) -- cycle;

       % NF
       \draw[-{latex},line width=2pt,color=purple] (-0.8,2.1) -- (-3,5);
       \draw[-{latex},line width=2pt,color=purple] (-3.05,5) -- (6,5);

       % tip
        \filldraw[color=gray!80] (-1,3) -- (-0.75,2.2) -- (-0.5,3) -- cycle;
        \draw[line width=5pt,color=gray] (-1.01,3) -- (-0.49,3) node[midway,above] {\textcolor{black}{Probe}}; 
        \draw[<->] (-0.35,2.5)--(-0.35,3) node[midway,right] {$\Omega$};

       % BS
        \draw[line width=2pt,color=black] (3.3536,-0.3536)--(3-0.3536,0.3536); %bottom
        \draw[line width=2pt,color=black] (3.3536,5+0.3536)--(3-0.3536,5-0.3536); %top
        \filldraw[] (3,5.75) node[] {BS};
        \filldraw[] (3,-0.75) node[] {BS};
        
        %Ref Mirrors
        \filldraw[color=black,fill=gray!50] (3-0.3536,2-0.3536)--(3+0.3536,2+0.3536)--(3+0.3536,3-0.3536) -- (3-0.3536,3+0.3536)--cycle;
        \filldraw[color=black,fill=gray!50] (4-0.3536,2-0.3536)--(4+0.3536,2+0.3536)--(4+0.3536,3-0.3536)--(4-0.3536,3+0.3536)--(5,3+0.3536)--(5,2-0.3536)--(4-0.3536,2-0.3536) node[midway,below] {RM};
        \draw[<->] (4-0.3536,3.5)--(5,3.5) node[midway,above] {$f$};
        
        %Detector
        \filldraw[color=black,fill=gray!50,very thick] (6,4.5)--(6,5.5)--(7,5)--(7,5.5)--(7,4.5)--(7,5)--cycle; 
        \draw[] (6.5,5.75) node {Detector};

        %PM
        \def\amov{0.15}
        \def\bmov{0.1}
        \filldraw[color=black,fill=gray!50] (0.5-0.3536-0.25+\amov,0+0.3536+\bmov)--(0.5-0.3536+\amov,0+0.3536+\bmov)..controls(0.5-0.175+\amov,-0.175+\bmov)..(0.5+0.3536+\amov,0.1-0.3536+\bmov)--(0.5+0.3536+\amov,0.1-0.3536-0.25+\bmov)--(0.5-0.3536-0.25+\amov,0.1-0.3536-0.25+\bmov)--cycle node[midway,left] {PM};

        \filldraw[color=black,fill=gray!50] (-3+0.3536+\amov,5+0.3536+0.25-\bmov)--(-3+0.3536+\amov,5+0.3536-\bmov)..controls(-3-0.175+\amov,5+0.175-\bmov)..(-3-0.3536+\amov,5-0.3536-\bmov)--(-3-0.3536-0.25+\amov,5-0.3536-\bmov)--(-3-0.3536-0.25+\amov,5+0.3536+0.25-\bmov)--cycle node[midway,above] {PM};

        \draw[<-] (0,-0.5) -- (0.5,-0.5);
        \draw[->] (0,-0.65) -- (0.5,-0.65);

        \draw[<-] (-3,0.8) -- (-2.5,0.8);
        \draw[->] (-3,0.65) -- (-2.5,0.65);
    \end{tikzpicture}
    \caption{Schematic illustration of the s-SNOM setup.}
    \label{fig:supp-sSNOM}
\end{figure}

\clearpage

\section{S5. Data processing details}
\label{sec:supp-data_processing}
Ideally, the complex-valued wavevector can be extracted directly from the Fourier spectrum. However, this requires infinitely long data, which is practically impossible. When the data is regularly sampled and of finite length $L$, then one can apply the classical Discrete Fourier Transform (DFT), which will have a resolution of $\Delta k = 2\pi/L$, corresponding to $\Delta N_m = \lambda/L \approx 0.04$ for our 40-$\upmu$m-long scans. Additionally, DFT of our scans will result in spectral leakage because of "windowing", since only a 40-$\upmu$m-long part of an otherwise infinitely long signal is transformed. The resolution can be improved by artificially increasing the length of the data by zeroes (so-called zero padding), but the Fourier spectrum will still be plagued by spectral leakage, resulting in spreading and overlapping of modes, which will reduce the efficiency of Fourier filtering (that is, the filtered mode in the direct space will contain a small amount of other modes, reducing the accuracy of the following fitting).

To mitigate the above issue of spectral leakage, we apply the Extended Discrete Fourier Transform (EDFT), developed by Liepins~\cite{liepins:EDFT}. Essentially, EDFT assumes a band-limited "true" spectrum (of unknown infinitely long data), and it iteratively extrapolates the measured data to minimize the difference between its DFT spectrum and the "true" spectrum. For a thorough summary of EDFT see Liepins~\cite{liepins:EDFT}. In our case, we applied EDFT with $5L$ extrapolated length (i.e., improving the resolution of DFT by 5 times) and 10 iterations (which is justified by observing no noticeable change in the Fourier spectrum after consecutive iterations). To minimize the influence of selected-by-hand parameters of the Fourier transform, we find the wavevector by the direct-space fitting of the Fourier-filtered mode. 

As mentioned in the main text, our measurements suffered from not perfect synchronization between the movement of the sample and the bottom parabolic mirror during the scan due to the lower quality of the mirror stage. This resulted in a small displacement ($\sim 100$ nm) of the incident spot relative to the flake edge. Since the spot size is relatively large (FWHM $\sim 3\,\upmu$m), this displacement has a negligible influence on the near-field amplitude, but it introduces phase wobbling.  To correct the phase ‘wobbling’ and determine $N_m$ without artificially lowering the estimated errors, we use the following two-step procedure (Figure~\ref{fig:supp-Flow_chart}):

\begin{figure}[!h]
    \centering
    \includegraphics[width=\linewidth]{  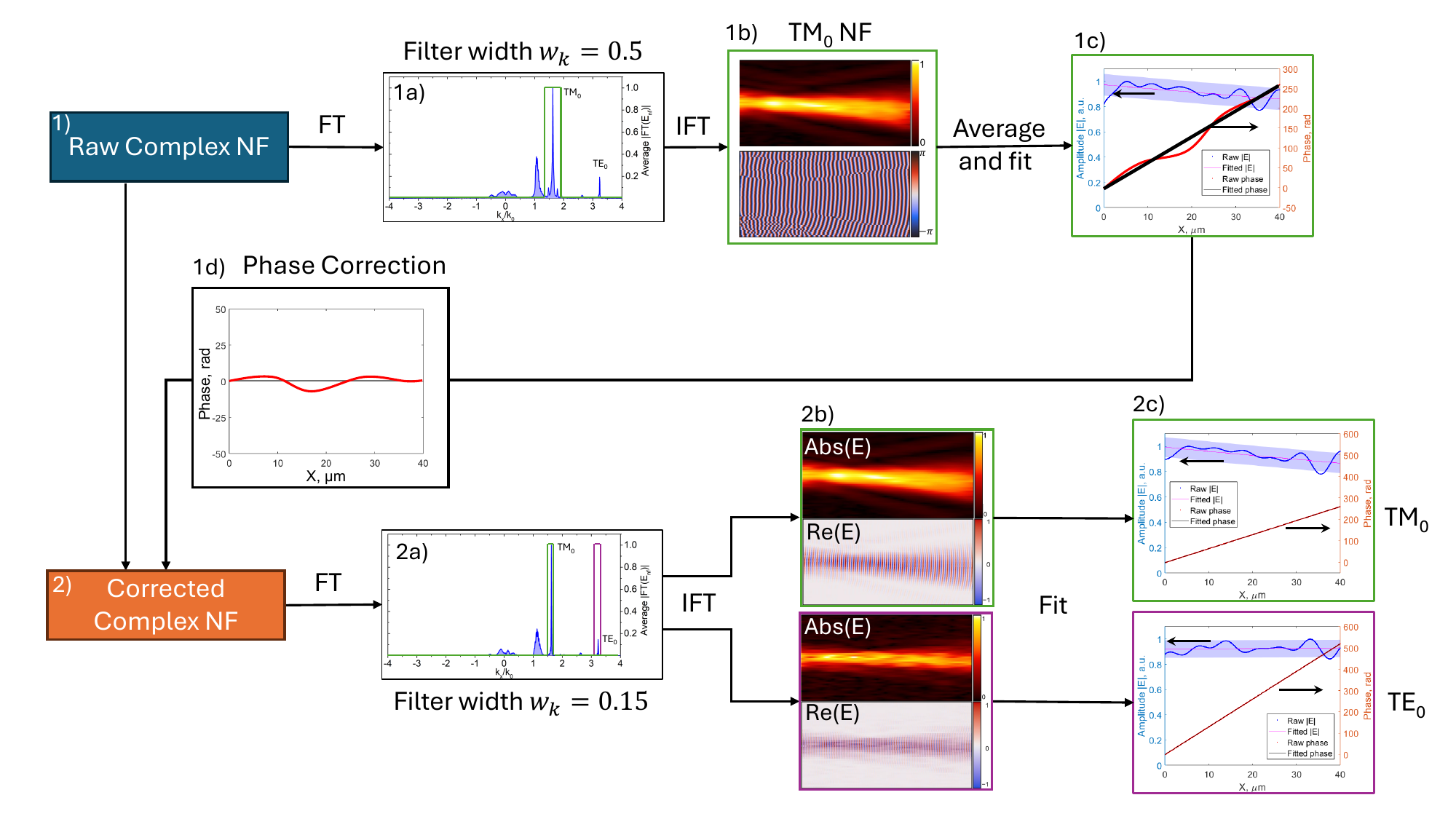}
    \caption{Near field data processing procedure to extract the effective mode indices. \textbf{1)} The raw complex near field map is Fourier transformed (using EDFT). \textbf{1a)} Filtering of the optical modes with a square window of width $0.5k_0$. \textbf{1b)} Followed by inverse Fourier transform. \textbf{1c)} The amplitude weighted unwrapped phase is linearly fitted. \textbf{1d)} The residuals of the linear phase fit is extracted and assumed as the phase correction. \textbf{2)} The correction phase is subtracted from the raw near-field phase. \textbf{2a-c)} Similarly to steps 1a-c, the corrected near field is Fourier transformed, each mode is filtered with a square window of the width $0.15k_0$, followed by fitting in the direct space.}
    \label{fig:supp-Flow_chart}
\end{figure}

1) First, we select the mode with the most prominent peak in the Fourier spectrum, which does not overlap with others, and filter it using a square window function with a width of $0.5 k_0$. Then it is inversely transformed back to real space, converted to 1D by performing integration along the $y$-direction, followed by a linear fit of the unwrapped phase. This process provides the residual phase, which is then subtracted from the raw data.

2) In the second iteration, the corrected complex near-field data is Fourier transformed again and filtered for each mode using a smaller square window function with a width of $0.15 k_0$, followed by the same procedures to provide fitted $N_m$. Importantly, to avoid artificial lowering of the uncertainty, we estimate the squared error of $N_m$ for each mode as the sum of the squared error in the second step and the squared error for the reference mode in the first step. 

\begin{figure}[!h]
    \centering
    \includegraphics[width=\linewidth]{  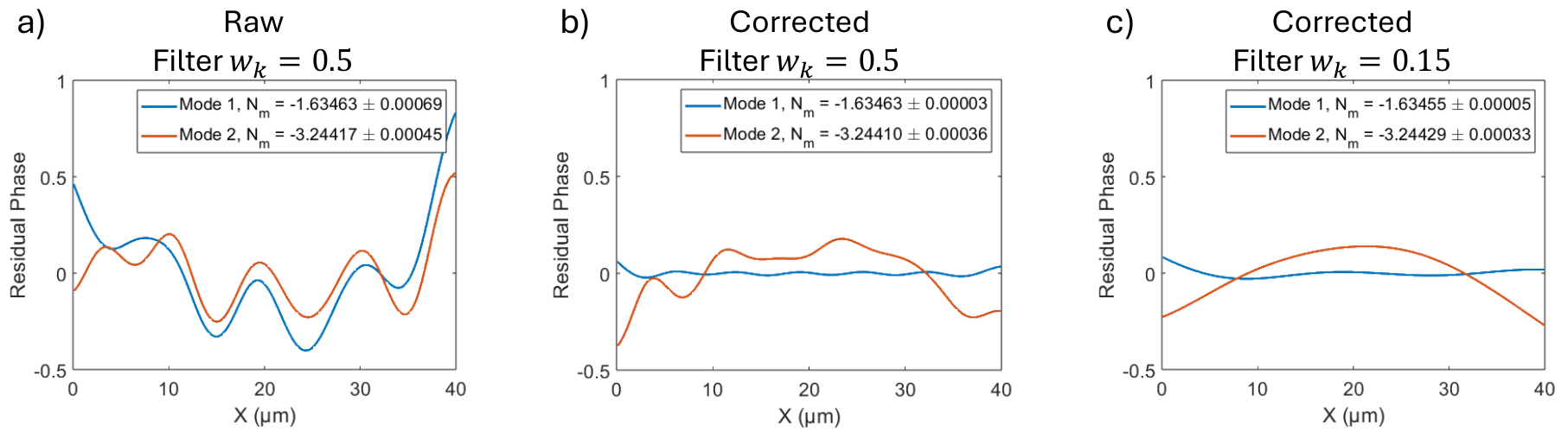}
    \caption{Residual phase along the scan direction of \textbf{a)} the raw data filtered for the two modes, \textbf{b)} after the correction using mode 1 and filtered using a square window with width $w_k=0.5k_0$, \textbf{c)} similarly as b) but with window width $w_k=0.15k_0$. Mode 1 corresponds to TM$_0$ and mode 2 to TE$_0$. Figure b) and c) does not show a large difference making the use of a the small window in the corrected data filtering step acceptable.
    }
    \label{fig:supp-Residual_phases}
\end{figure}

To justify that the residual phase is indeed a phase correction for the whole data, we compare residual phases for both modes, which demonstrates strong correlation (Figure~\ref{fig:supp-Residual_phases}a). As a result, correcting the phase from the residual phase of the first mode not only reduces the residual phase of the same mode (as expected), but also reduces the residual phase of the second mode (Figure~\ref{fig:supp-Residual_phases}b). 
The difference in using a window width of $0.5k_0$ and $0.15k_0$, as illustrated in Figure~\ref{fig:supp-Residual_phases}b-c, does not change the estimated effective index and its error significantly. The smaller filtering window of $0.15k_0$ was chosen to separate closely spaced TE$_1$ and TM$_0$ modes of the flake C (see Figure~3 in the main text).

\begin{figure}[!h]
    \centering
    \includegraphics[width=0.75\linewidth]{  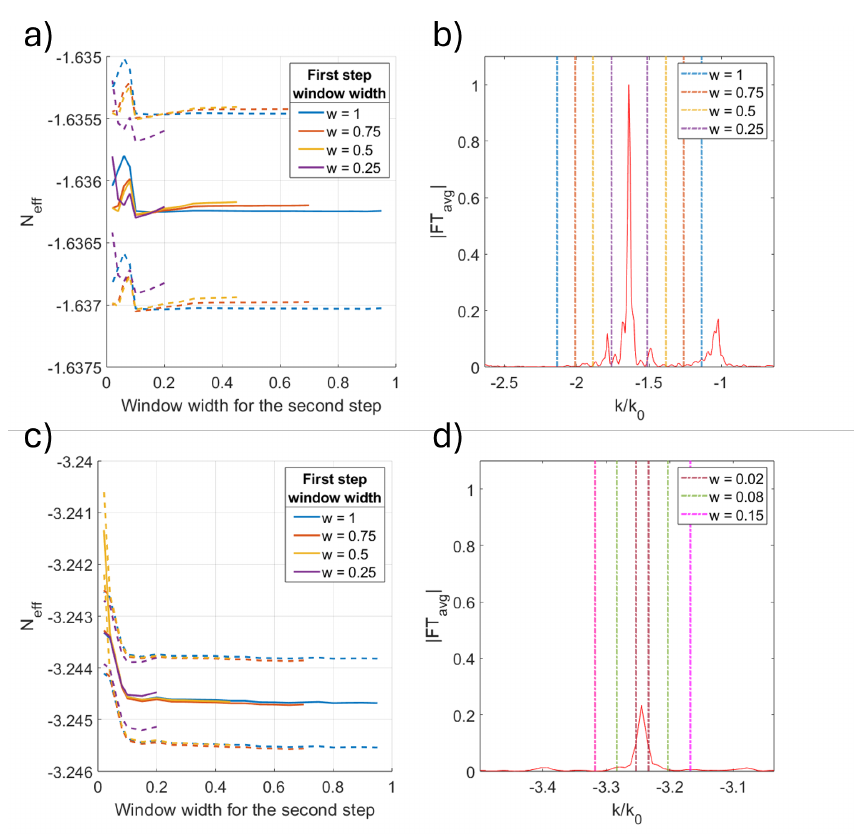}
    \caption{\textbf{a)} and \textbf{c)} Filter convergence for the TM$_0$ and TE$_0$ modes respectively and the two steps for a measurement on the 185 nm flake (solid lines). The dashed lines indicate the error bounds. \textbf{b)} Momentum spectra (solid line) around the TM$_0$ mode with the filter widths (dash-dot lines) for the first step and \textbf{d)} for the second step for the TE$_0$ mode.}
    \label{fig:Filter_test}
\end{figure}

When applying a square window function (filter), it is crucial to consider the impact of the selected window width. A window width that is too narrow may exclude valuable information, while an excessively wide window may encompass additional peaks in the spectrum, thereby reducing accuracy. In the initial step, a width of $0.5k_0$ is used to ensure a proper phase correction. However, varying this width will result in slight changes in the outcomes of the subsequent step. Figure~\ref{fig:Filter_test} illustrates this effect for the TM$_0$ and TE$_0$ modes on the 185 nm thick flake B. It is evident that altering the initial width causes minor variations in the fitted effective mode index; however, these are negligible compared to the inherent measurement errors. Even when the width is adjusted in the second step, the results remain consistent unless the window width approaches the peak width, causing asymmetry and skewing the effective mode index. In our data processing procedure, we utilize a width of $0.15k_0$ in the second step (as mentioned above), which is the smallest feasible width that does not significantly impact the results and is still able to differentiate modes that are close in momentum space.

\clearpage

\section{S6. Fitting the effective mode index data to extract the permittivity}
\label{sec:supp-epsi_fitting}
There are multiple ways to determine the permittivity from the measurements, each varying in complexity and accuracy. The most straightforward approach involves using a couple of effective mode indices (one for TE and one for TM mode) measured on a flake with a thickness $t$ and applying them directly in the dispersion relations. First, $\varepsilon_\parallel$ is determined directly from the dispersion equation for TE modes (see eq. 2a the main text), which is then used in the equation for TM modes to find $\varepsilon_\perp$ (2b in the main text). However, this method assumes a one-way dependence between the permittivity components and considers only a single flake thickness $t$ and a couple of effective indices. Moreover, it completely ignores the inaccuracy of $t$.

To account for all measurements and estimated errors, a least-squares approach is used, where estimated errors are used as weights to ensure the proper contribution from each measurement. In our approach we are looking for an unknown MoS$_2$ anisotropic permittivity $\varepsilon^\mathrm{fit}$ and a set of exact flake thicknesses $\mathbf{t^\mathrm{fit}}$, which will minimize the following normalized deviations between measured and exact (or calculated) parameters: 
\begin{equation}
   \mathbf{F}(\varepsilon^\mathrm{fit},\mathbf{t}^\mathrm{fit})=\sum_i\left[
   \left(\frac{t^\mathrm{fit}_i-t_i}{\Delta t_i}\right)^2 + 
   \sum_m\left(\frac{N(\varepsilon^\mathrm{fit},t^\mathrm{fit}_i)-N_{i,m}}{\Delta N_{i,m}}\right)^2
   \right],
   \label{eq:squares}
\end{equation}
where the summation goes over all flakes (labeled with index $i$) and all supported TE and TM modes (labeled with index $m$), whose effective mode indices are calculated from $\varepsilon$ and $\mathbf{t}^\mathrm{fit}$ by using the dispersion equations.

The above least-squares approach can be solved directly by searching all unknown 8 parameters at once (2 parameters of $\varepsilon$ and 6 parameters of $t^\mathrm{fit}$ for flakes A-F). However, such fitting is relatively heavy and might be unstable, therefore we decided to apply the following nested fitting:

1) If $\varepsilon$ is known, then the thickness of each flake $t^\mathrm{fit}_i$ can be determined by the least-squares fitting, minimizing the following corresponding deviation:
\begin{equation}
   \mathbf{F_i}(t^\mathrm{fit}_i)=
   \left(\frac{t^\mathrm{fit}_i-t_i}{\Delta t_i}\right)^2 + 
   \sum_m\left(\frac{N(\varepsilon,t^\mathrm{fit}_i)-N_{i,m}}{\Delta N_{i,m}}\right)^2,
\end{equation}
which implicitly defines thickness as a function of $\varepsilon$, $t^\mathrm{fit}_i\left(\varepsilon\right)$. Unlike in the straightforward approach, here it is only one fitting parameter $t^\mathrm{fit}_i$.

2) The above implicit definition of $t^\mathrm{fit}_i$ is used in Eq.~\ref{eq:squares}, leaving only 2 fitting parameters: $\varepsilon_\parallel$ and $\varepsilon_\perp$.

Thus, our nested fitting procedure implicitly defines the permittivity and flake thicknesses as a function of measured parameters:
\begin{equation}
   (\varepsilon^\mathrm{fit},\mathbf{t}^\mathrm{fit})=Q\left(\mathbf{t},\mathbf{N},\Delta\mathbf{t}, \Delta\mathbf{N}
   \right).
\end{equation}

We have used a function 'fsolve' in MATLAB software to find effective mode indices from dispersion equations and do the least-squares fitting. The initial guess for the fitting parameters are the corresponding flake thicknesses measured by AFM and the permittivity taken from Ermolaev et al.~\cite{Ermolaev:2021}  ($\varepsilon_\parallel=16.56$, $\varepsilon_\perp=6.43$ at our free-space wavelength of $1570$ nm).

To estimate the uncertainty of our fitted parameters, we have assumed all our measurements being independent (resulting in the upper bound estimation of the uncertainty). By using standard sensitivity analysis, we estimate the error contribution of each measured value as its derivative, multiplied by the corresponding error. Then all these contributions are squared and added to estimate the uncertainty of the permittivity:

\begin{equation}
    \Delta\varepsilon=\sqrt{\sum_i\left[
    \left(\frac{\partial Q}{\partial t_i}\Delta t_i\right)^2 + 
    \sum_{m}\left(\frac{\partial Q}{\partial N_{i,m}}\Delta N_{i,m}\right)^2
    \right]}.
\end{equation}

The partial derivatives of $Q$ are obtained simply by a finite difference approximation:
\begin{equation}
    \frac{\partial Q}{\partial t_j}=\frac{Q\left(\mathbf{t}(t_j+\Delta_t),\mathbf{N}\right)-Q(\mathbf{t},\mathbf{N})}{\Delta_t},
\end{equation}
\begin{equation}
    \frac{\partial Q}{\partial N_{j,i}}=\frac{Q\left(\mathbf{t},\mathbf{N}(N_{j,i}+\Delta_N)\right)-Q(\mathbf{t},\mathbf{N})}{\Delta_N},
\end{equation}
where $\Delta$ is a small perturbation.
The changes in $\Delta_t$ and $\Delta_N$ are consistent across all derivatives and should be selected to be as small as possible, while still being sufficiently large to avoid issues related to numerical precision. We chose $\Delta_t=1$ nm and $\Delta_N=0.001$, which corresponds approximately to $10\%$ of the smallest error in $\Delta \mathbf{t}$ and $\Delta \mathbf{N}$.

\end{document}